\documentclass{aa}
\usepackage{graphicx}
\usepackage[varg]{txfonts}
\usepackage{longtable}
\usepackage{color}

\begin{document}

   \title{Radial-velocity variations due to meridional flows in the Sun and solar-type stars: impact on exoplanet detectability}


   \titlerunning{Radial velocity variations due to meridional flows in the Sun and solar-type stars}

   \author{N. Meunier \inst{1}, A.-M. Lagrange \inst{1}
          }
   \authorrunning{Meunier et al.}

   \institute{
  Univ. Grenoble Alpes, CNRS, IPAG, F-38000 Grenoble, France\\
\email{nadege.meunier@univ-grenoble-alpes.fr}
             }

\offprints{N. Meunier}

   \date{Received ; Accepted}

\abstract{Stellar variability due to magnetic activity and flows at different spatial scales  strongly impacts radial velocities. This variability is seen as oscillations, granulation, supergranulation, and meridional flows. The effect of this latter process is currently poorly known but could affect exoplanet detectability.}
{We aim to quantify the amplitude of the meridional flow integrated over the disc and its temporal variability, first for the Sun, as seen with different inclinations, and then for other solar-type stars. We then want to compare these amplitudes with low-mass exoplanetary amplitudes in radial velocity.}
{We used long time series (covering two 11-year cycles) of solar latitudinal meridional circulation to reconstruct its integrated contribution and study its properties. We then used scaling laws from hydrodynamical simulations relating the amplitude of the meridional flow variability with stellar mass and rotation rate to estimate the typical amplitude expected for other solar-type stars. }
{We find typical rms of the order of 0.5-0.7 m/s (edge-on) and 1.2-1.7 m/s (pole-on) for the Sun (peak-to-peak amplitudes are typically 1-1.4 m/s and 2.3-3.3 m/s resp.), with a minimal jitter for an inclination of 45-55$^\circ$. This signal  is significant compared to other stellar activity contributions and is much larger than the radial-velocity signal of the Earth. The variability is strongly related to the activity cycle, with maximum flows during the descending phase of the cycle, and possible variability on timescales lower than the cycle period. Extension to other solar-type stars shows that the variability due to meridional flows is dominated by the amplitude of the cycle of those stars (compared with mass and rotation rate), and that the peak-to-peak amplitudes can reach 4 m/s for the most variable stars when seen pole-on. The meridional flow contribution sometimes represents a high fraction of the convective blueshift inhibition signal, especially for  quiet, 
 low-mass stars. For fast-rotating stars, the presence of multi-cellular patterns should significantly  decrease the meridional flow contribution to the radial-velocity signal.  }
{Our study shows that these meridional flows could be critical for exoplanet detection. Low inclinations are more impacted than edge-on configurations, but these latter still exhibit significant variability. Meridional flows also degrade the correlation between radial velocities due to convective blueshift inhibition and chromospheric activity indicators. This will make the correction from this signal challenging for stars with no multi-cellular patterns, such as the Sun for example, although there may be some configurations for which the line shape variations may be used if the precision is sufficient. }


\maketitle

\section{Introduction}

Stellar variability strongly  impacts exoplanet radial-velocity (RV) detection and mass characterisation in different ways. Firstly, it is affected by stellar magnetic activity  \cite[e.g.][]{saar97,hatzes02,saar03,wright05,desort07,lagrange10b,meunier10a,boisse12,dumusque14,borgniet15,dumusque16}. Small-scale dynamics such as oscillations, granulation, and supergranulation also affect RVs \cite[][]{dumusque11,cegla13,meunier15,cegla18,meunier19e,cegla19}. All of these effects have been studied. Much less attention has been paid to the effect of large-scale flows such as meridional circulation. Meridional flows, as in differential rotation, arise from the redistribution of angular momentum caused by turbulence. The only  estimation of integrated RV that we are aware of was performed by \cite{makarov10b}. This latter author simulated the solar RV over a cycle, including magnetic activity (and in particular the convective blueshift inhibition in plages) and meridional flows. \cite{makarov10b} implemented this simulation for the Sun seen edge-on only, and the amplitude of the contribution of the meridional flows is not entirely clear since it was superposed to the other contributions. He derived a total amplitude (including convective blueshift inhibition) of about 7 m/s. We expect that the contribution of the meridional flows he simulated  is lower than 1-2 m/s assuming the full amplitude he found (7 m/s) is compatible with the results of \cite{meunier10}, who did not take meridional flows into account.  

Here, we aim to quantify the amplitude and timescale of the meridional flow contribution to the RV signal, and to compare the former with the amplitude of an Earth-like planet. 
We also expect a reversal at a certain inclination because the meridional flows are predominantly poleward, and therefore taking stellar inclination into account is crucial. Long time series of high-quality solar measurements can be used to estimate the effect of meridional flows for the Sun seen with different inclinations. In addition, we use recent results from 3D hydrodynamical simulations to extrapolate the solar values to stars other than the Sun with different spectral types and activity levels. 

The outline of this paper is as follows. We first consider the solar case in Sect.~2 for various inclinations and solar latitudinal profiles. The resulting RV contribution is compared to a planetary signal in the habitable zone and other solar-activity contributions. These results are then extrapolated to other solar-type stars in Sect.~3, and we conclude in Sect.~4.

\section{The solar case}

We use published solar meridional flow profiles versus latitude and time to build the integrated meridional circulation (MC) at each time step, mc(t), for different solar inclinations. We first describe  the latitudinal profiles used as input and then describe the procedure used to compute the disc integration and the results. 

\subsection{Input observed meridional circulation}

There is a long history of solar MC measurements using various techniques \cite[e.g.][]{duvall79,labonte82,howard86,ulrich88,komm93,snodgrass96,hathaway96,nesmeribes97,meunier99}, with predominantly poleward flows. \cite{meunier99} observed weaker flows during the ascending phase of the cycle compared to solar minimum, already seen by \cite{komm93} with larger uncertainties, as well as a converging flow pattern around the activity belt which is not present during the minimum. This pattern was later observed in other studies using different techniques \cite[e.g.][]{zhao04,lin18}, while some diverging patterns have also been observed \cite[][]{snodgrass96,chou01}.

Here, we select monitorings of MC  with good temporal coverage of typically about a solar cycle or more so as to have a complete view of its temporal variability along the cycle (see Table~\ref{tab_mc}). We focus on the time series obtained by \cite{ulrich10} because
MC derived from full-disc Dopplergrams is the most interesting for our purposes, as it corresponds to flows formed at an altitude in the photosphere that is very similar to the altitude we are interested in when observing stars in the optical. The different time series obtained with the other inputs are described and compared  in Appendix A to show the robustness of the results. Another relatively long time series by \cite{gizon10} is available but we do not include it here because it is slightly shorter than the others and is the combination of two  short time series obtained with different techniques.

\begin{table*}
\caption{Input meridional circulation latitudinal profiles}
\label{tab_mc}
\begin{center}
\renewcommand{\footnoterule}{}  
\begin{tabular}{lllll}
\hline
        Reference & Name & Temporal coverage & Nb of points & Analysis  \\
\hline
        \cite{ulrich10} & U10 & 1986-2009 & 24  & Dopplergrams\\
        \cite{basu10} & BA10 & 1996-2009 & 15 & Ring diagram\\
        \cite{hathaway10} & HR10 & 1996-2009 & 166 & Magnetic feature tracking\\
        \cite{hathaway11} & HR11 & 1996-2010 & 180 & Magnetic feature tracking\\
\hline
\end{tabular}
\end{center}
\end{table*}

\begin{figure} 
\includegraphics{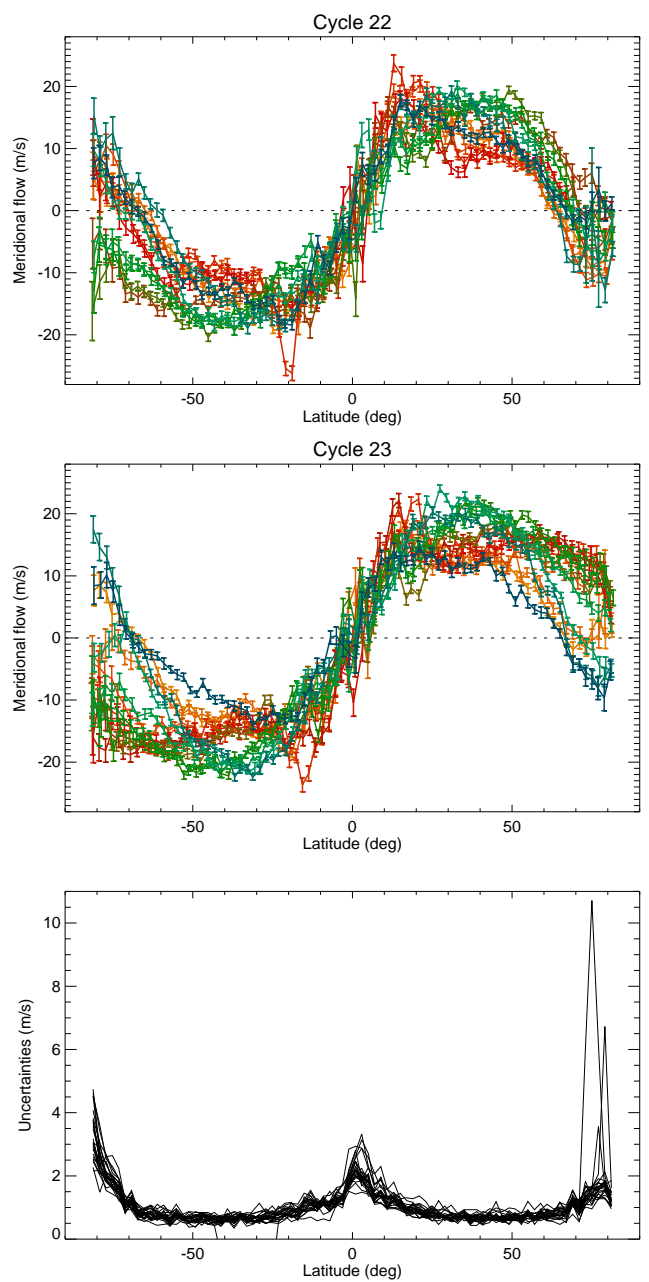}
\caption{
        Latitudinal profiles of meridional flows from U10 for cycles 22 (upper panel) and 23 (middle panel). Positive flows are northwards. The colour code indicates time, from orange (beginning of each cycle) to blue (end of each cycle), equally spread in cycle phase. The lower panel shows the uncertainties vs. latitude (one curve per year).
}
\label{lat}
\end{figure}

\cite{ulrich10} (hereafter U10) measured the meridional flows with a very good temporal coverage (1986-2009, i.e. covering two solar cycles and including three activity minima) from Dopplergrams obtained at the Mount Wilson Observatory. Each yearly latitudinal profile has been digitised (the uncertainty on this numerisation is much smaller than the error bars on the meridional flow estimates) and is shown in Fig.~\ref{lat}. 
Some profiles show sign reversals at high latitudes, indicating the presence of a second cell in the circulation. Ulrich (2010)argues that they are of solar origin because he took effects such as the limb convective blueshift into account in the analysis, although this was discussed by \cite{hathaway12} who casts some doubt on the reality of the flows. Differences between the results of these latter two authors could be due to the fact that they use different methods probing different layers. Small features across the equator are also present for a few years, but they are not likely to be of solar origin: U10 argues that these could be an artefact coming from the data processing (lack of equatorial symmetry at higher latitude). Uncertainties on the latitudinal profiles are typically around 0.5-1 m/s at medium latitudes and can reach several metres per second at high latitudes (and up to 2 m/s close to the equator).

\subsection{Building of the integrated meridional circulation}

\begin{figure} 
\includegraphics{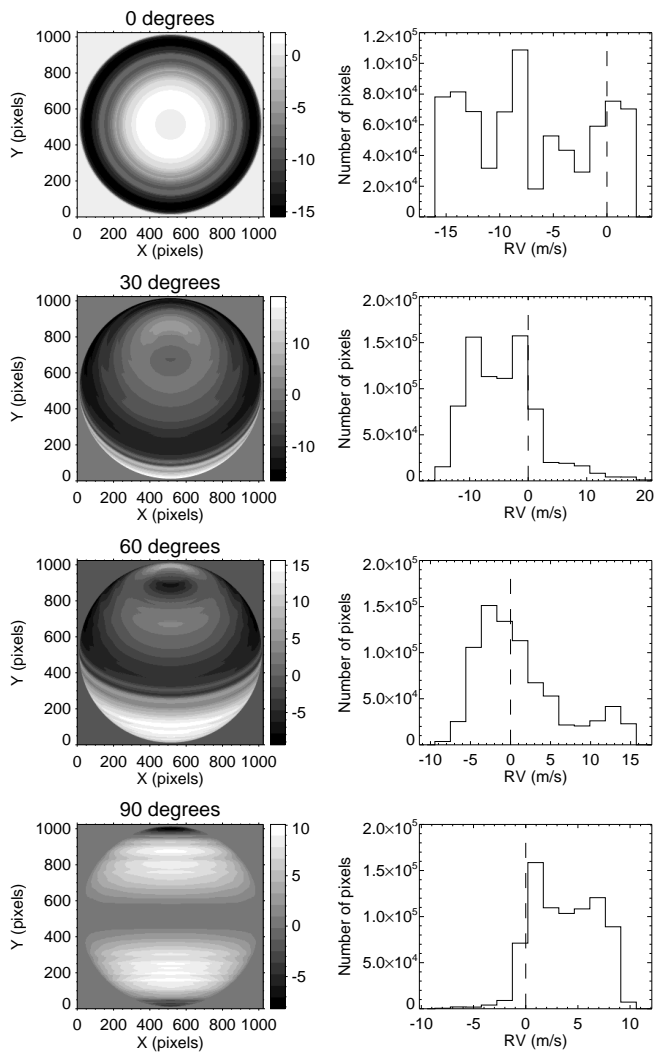}
\caption{
        Examples of maps (left panels) and corresponding velocity distributions (right panels) over the solar disc for inclinations of 0$^\circ$, 30$^\circ$, 60$^\circ$, and 90$^\circ$ (from top to bottom), for the 1986 latitudinal profile of Ulrich (2010). The levels are in metres per second, and are positive for redshifts. }
\label{exmap}
\end{figure}

We subsequently consider the MC latitudinal profile at a given time step. We build a map of the flows over the solar disc according to this profile, taking projection effects into account, for a given solar inclination. Figure~\ref{exmap} shows examples of such maps for different inclinations, together with the velocity distributions, which have a complex shape. We then integrate the flows over the disc, after weighting them with the limb-darkening function of \cite{claret03}. The resulting integrated mc(t) is then obtained for inclinations between 0$^\circ$ (pole-on configuration) and 90$^\circ$ (edge-on configuration), with a step of 5$^\circ$, for each date. 
We do not weight the flows with the difference in flux in spots and plages because the impact should be small, especially as we consider long-term variability; in addition, they are located in converging flow regions which should partially compensate each other, hence the lower contribution to the integrated MC variability from these areas. 

The estimation of solar meridional circulation is usually not possible at high latitude because of strong projection effects. The maximum latitude depends on the technique but is typically in the 50-80$^\circ$ range. For higher latitudes, we interpolated the published flows between the MC value at the highest latitude and a MC value of zero at the pole (separately in each hemisphere). This is justified by the necessity to have a single turning point for the flows, and by the results of \cite{hathaway12} who observed significant meridional flows up to the poles.

In addition to the integrated meridional flows mc(t), we study the line shape variability over time. For this purpose, we attribute to each pixel of the disc a spectrum consisting in one spectral line which is shifted according to the projected MC flow at this pixel and projected rotational velocity. The spectral lines from all pixels are then combined (after weighting by the limb-darkening function). We use the synthetic solar optical spectra used in the SAFIR software \cite[][]{galland05} from \cite{kurucz93}, and use only a single line here for simplification, at 5300~\AA, i.e. in approximately  the middle of the HARPS spectrograph wavelength coverage. We then compute the bisector of this line after integration over the disc to check whether meridional flows leave a signature in the line shape. 
To estimate the variation of the bisector shape, we compute the variable BIS at each time
step, which is the difference between two quantities, $\lambda_1$ and $\lambda_2$, after conversion to a velocity. These quantities are computed as follows. We use $I$ to denote the intensity at the line core and $I'$ the line depth (1-$I$). Values of $\lambda_1$ and $\lambda_2$ are then the wavelengths of the middle of the line profile for intensities between $I$ and $I$+0.2$I'$ and between $I$+0.7$I'$ and $I$+0.9$I'$ respectively. 

Finally, the time series will be compared to two activity indicators: the spot number from SIDC (Solar Influences Data analysis Center) and the S-index from the Sacramento Peak Observatory. The objective is to compare the phase of the different series during the cycle.

\subsection{Results}
 
In this section, we first present the time series obtained from this reconstruction. We characterise the amplitude of the variability for different inclinations. Finally, we compare the expected MC signal with an Earth-like signal in the habitable zone and with the contribution due to stellar  magnetic activity and other stellar flows. 

\begin{figure} 
\includegraphics{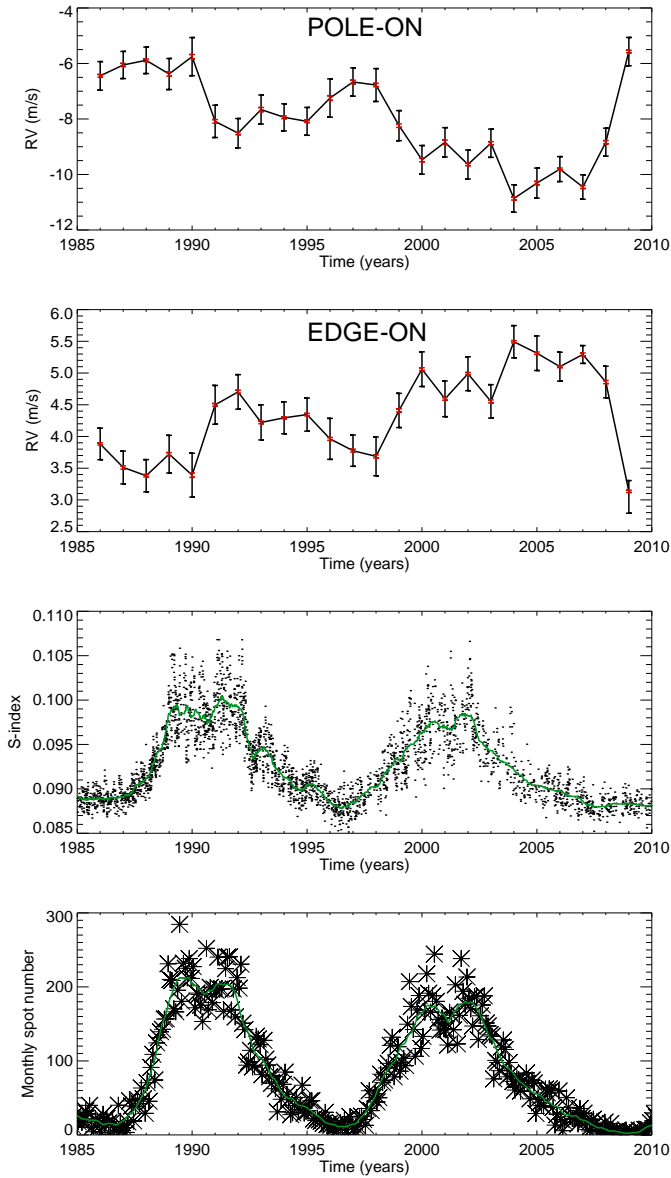}
\caption{
Time series of integrated meridional flows for pole-on configuration (first panel) and edge-on configuration (second panel) for U10 reconstruction. Two types of uncertainties are shown (see text): a lower limit (red) and an upper limit (black). The third panel shows the S-index from Sacramento Peak Observatory between 1985 and 2011 (the solid line corresponds to a smoothing over 100 days). The last panel shows the monthly spot number (stars), superposed on a smoothed time series as a solid line.
}
\label{temps1}
\end{figure}

\begin{figure} 
\includegraphics{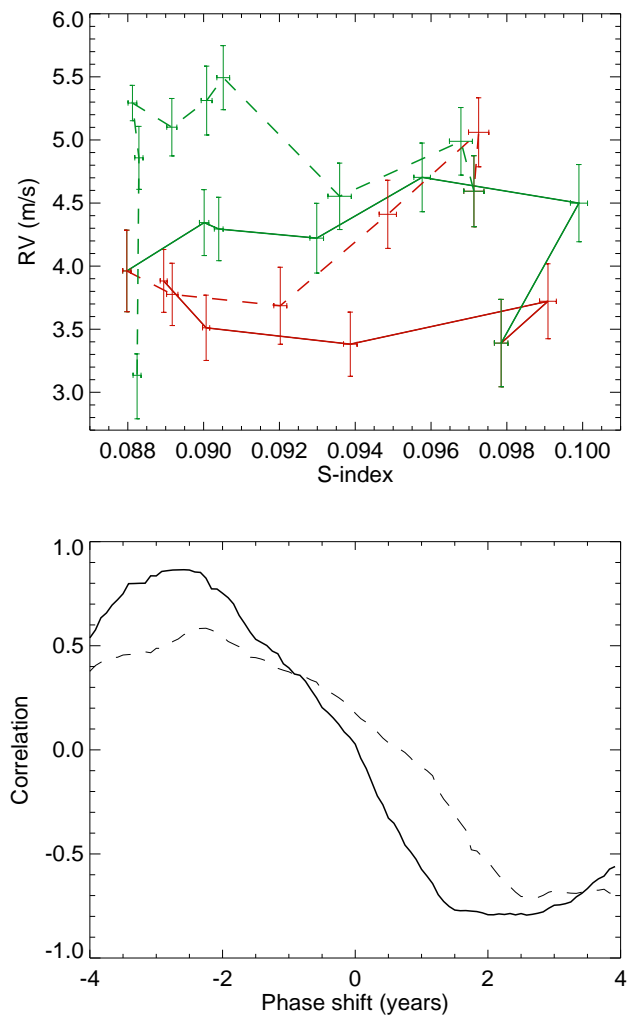}
\caption{
        Meridional flow vs. S-index (upper panel) for pole-on configuration for cycle 22 (solid line) and cycle 23 (dashed line): ascending phases are shown in red and descending phases in green. The uncertainties in RV are the upper limits. The lower panel shows the cross-correlation functions between S-index and meridional flows for cycles 22 (solid line) and 23 (dashed line).  
}
\label{hyst1}
\end{figure}

\begin{figure} 
\includegraphics{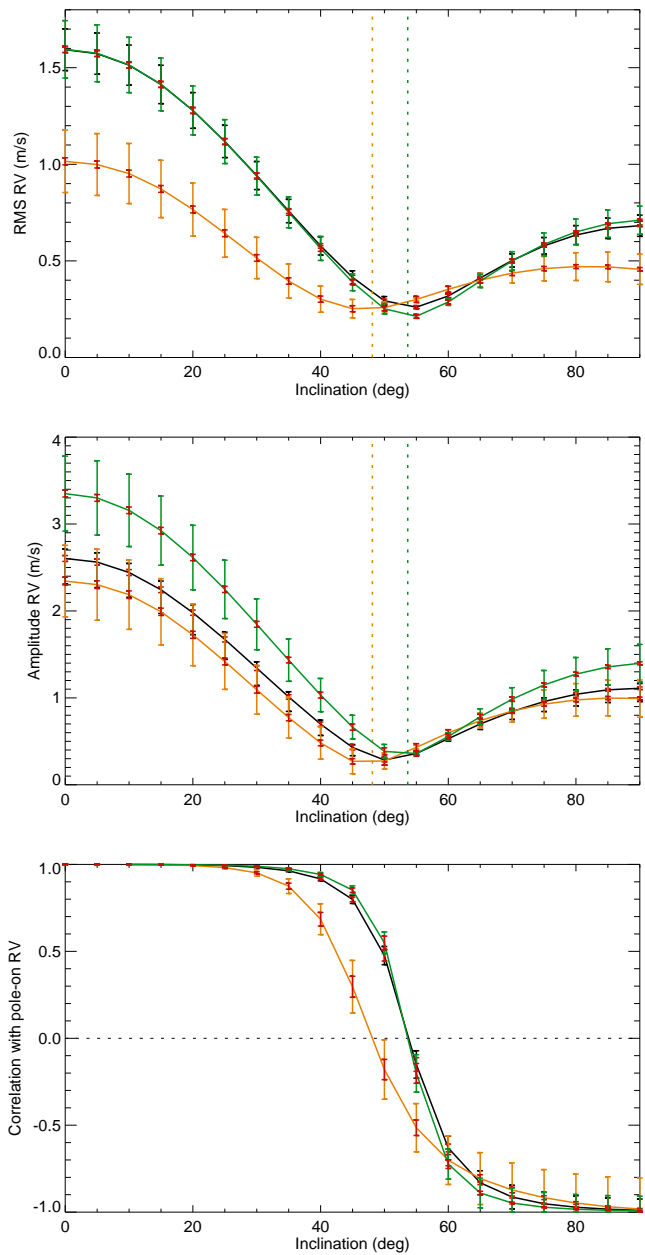}
\caption{
Root mean square of meridional flow time series (upper panel), peak-to-peak amplitude (middle panel), and correlation of meridional time series with pole-on time series (lower panel) vs. stellar inclination, for U10 reconstruction: full time series (black), cycle 22 (orange), and cycle 23 (green). The uncertainties with the same colour correspond to the upper limits, and those in red to the lower limit. The vertical dotted lines indicate the position of the reversals for the two cycles. 
}
\label{rms}
\end{figure}

\begin{figure} 
\includegraphics{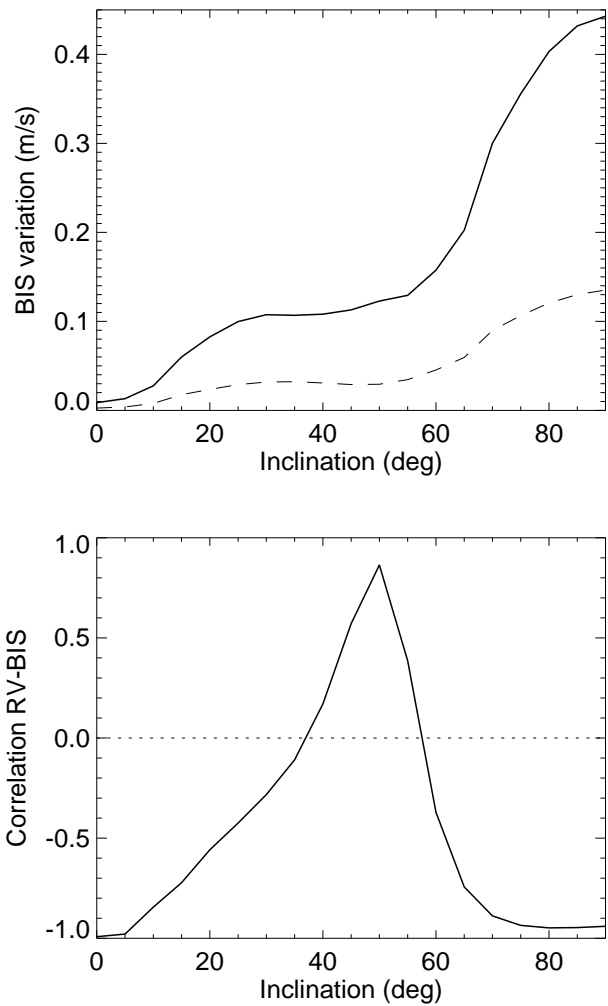}
\caption{
        Peak-to-peak amplitude (solid line) and rms of BIS (dashed line) vs. inclination (upper panel), and correlation between BIS and mc(t) (lower panel), to be compared with rms variation in Fig.~\ref{rms}. } 
        \label{bis}
\end{figure}

\begin{table*}
\caption{Solar rms and amplitude of integrated meridional flows}
\label{tab_rms}
\begin{center}
\renewcommand{\footnoterule}{}  
\begin{tabular}{lllll}
\hline
        MC  & RMS & RMS & $\Delta$ & $\Delta$ \\
time     & edge-on & pole-on & edge-on & pole-on \\
series      & (m/s) & (m/s) &  (m/s) & (m/s)  \\
\hline
{\bf U10 cycle 22}      & {\bf 0.46 (0.01;0.07)} & {\bf 1.02 (0.02;0.17)} & {\bf 0.50 (0.01;0.21)}  & {\bf 1.17(0.01;0.39)} \\
{\bf U10 cycle 23}      & {\bf 0.71 (0.02;0.08)} & {\bf 1.60 (0.02;0.14)} & {\bf 0.70 (0.01;0.20)}  & {\bf 1.67 (0.20;0.38)} \\
BA10 cycle 23   & 0.80 (0.05;0.06)& 1.97 (0.10;0.10) & 0.98 (0.07;0.19) & 2.39 (0.14;0.29) \\
HR11 cycle 23 & 0.59 (0.02;0.02)& 1.11 (0.03;0.04) & 0.64 (0.03;0.07)& 1.15 (0.06;0.11) \\
\hline
\end{tabular}
\end{center}
\tablefoot{The rms is computed on each time series, which have different durations depending on input latitudinal profiles. $\Delta$, also in m/s, is the amplitude of a sinusoidal fit (and 2$\Delta$ the peak-to-peak amplitude). Values between parentheses represent the low and upper uncertainties (see text). } 
\end{table*}

\subsubsection{Time series}

The time series for input latitudinal profiles from U10 and two configurations (edge-on and pole-on) are shown in Fig.~\ref{temps1}. The integrated meridional flow is on average negative when the star is seen pole-on, and positive when it is seen edge-on, because the flows are predominantly poleward, with an average of respectively -8.0 m/s and 4.3 m/s.  The MC time series are globally linked  to the cycle, but with a complex relationship: from Fig.~\ref{temps1}, the comparison of the pole-on time series with activity indicators shows that the meridional circulation is lower at cycle maximum compared to cycle minimum; furthermore, the meridional circulation (in absolute value) is larger during the descending phase of the cycle, as illustrated in Fig.~\ref{hyst1}. There is also a significant difference between the two cycles: for cycle 22, the maximum is closer to the maximum of the cycle, while for cycle 23 the maximum is closer to the end of the cycle and their amplitudes appear to be different. 
We attempt to estimate the shift between the activity cycle and MC (derived from U10) by computing the cross-correlation between the S-index and mc(t) shown in the lower panel of Fig.~\ref{hyst1}: the maximum of the correlation and the minimum are observed for lags different from zero; however the distance between the two peaks is much lower than the cycle period, meaning that a possible time lag is ill-defined.

The uncertainties on each mc(t) value are computed using two assumptions, providing upper and lower limits. The procedure is described in Appendix B.
It is likely that the true uncertainties lie between these two estimates, and both are shown in Fig.~\ref{temps1}. Even when considering the upper uncertainties, we conclude that the long-term variation (cycle) in Fig.~\ref{temps1} is significant, while the variability on smaller timescales may not be. 
The rms (root-mean-square) of each time series for all input latitudinal profiles is shown in Table~\ref{tab_rms}. The most interesting time series (i.e. from U10) has rms values between 0.46 and 0.71 m/s (edge-on). They are higher for the pole-on configuration (1.02-1.59 m/s). We note that although cycles 22 and 23 are of similar amplitude (from the chromospheric or spot number point of view), the variability of the RV due to meridional flows is larger by about 50\% during cycle 23, meaning the relationship between amplitude of the integrated meridional component and the cycle amplitude is complex, because this difference is significant. The amplitude $\Delta$ is the amplitude of the sinusoidal fit, and the corresponding peak-to-peak amplitude, 2$\Delta$, is used in most of the paper. We conclude that using other latitudinal profiles for MC gives RV amplitudes of similar magnitude, although they are either slightly larger (BA10) or slightly smaller (HR10, not shown here because very similar to HR11, and HR11).

An important question is whether the variability on timescales smaller than the cycle period is real. From a theoretical point of view,  numerical magneto-hydrodynamical simulations suggest variability  on long timescales only (A. Strugarek, private communication), although short-timescale variability is present due to the activity pattern (see e.g. the two lower panels of Fig.~\ref{temps1}). Such variability has been reported  in the past  \cite[][]{hathaway96} from the latitudinal profiles, and our reconstructions shown in Appendix A also seem to exhibit some year-to-year variability  as well as below 1 year for the reconstructions from HR10 and HR11. However, a first issue is the lack of correspondence  between the different reconstructions on those timescales (except between HR10 and HR11 since they correspond to very similar  analyses), which casts some doubt on the reality of such variability. A second difficulty arises from the uncertainties on the integrated RV: in Appendix B, we estimate the errors based on two extreme assumptions (correlated and uncorrelated between latitudinal bins), and we estimate that the true uncertainties lie between the two. Unfortunately, the lower bound uncertainties are compatible with a significant  short-timescale variability while the upper bound uncertainties are not. In conclusion, we cannot conclude whether the variability at short timescales is significant.

\subsubsection{Dependence on stellar inclination}

The rms over each time series is computed for all inclinations and is shown in Fig.~\ref{rms} for the two cycles available from U10. The rms is maximum for the pole-on configuration (0$^\circ$) and is minimal (below 0.3 m/s) around 45$^\circ$ for cycle 22 and 55$^\circ$ for cycle 23, which also corresponds to the reversal shown in the lower panel. Even with the upper limit for the uncertainties, the differences between the two cycles are significant. Slightly different shapes of the latitudinal  profiles are likely to be responsible for the difference in the reversal inclination. 
The second panel shows the peak-to-peak amplitudes, with similar conclusions about the reversal for inclinations around 50$^\circ$, the difference between the cycles, and the difference between the pole-on and edge-on configurations.  The reversal in sign is illustrated on the lower panel of  Fig.~\ref{rms}, which shows the correlation between the meridional time series at different inclinations with the pole-on time series. 

\subsubsection{Effect on the line shape}

The bisector span (BIS; see definition and computation method in Sect.~2.2), computed without noise on the spectra in these simulations, varies over time.
The results are shown in Fig.~\ref{bis}. 
The correlation between BIS and RV is close to -1 for pole-on (-0.99) and edge-on (-0.94) configurations but becomes positive for medium inclinations, with a maximum of 0.86 for 50$^\circ$. The rms of the BIS increases from pole-on to edge-on with values up to 0.14 m/s. The peak-to-peak amplitude is small in general, but can reach up to 0.4 m/s for edge-on configurations, which may be detectable in some cases.  This means that MC will not significantly affect this observable in many cases and the BIS cannot be used to correct the RV signal for the meridional flows for low and intermediate inclinations. This is probably due to the fact that, unlike spot or plage crossing across the disc, meridional flow maps are symmetric with respect to the central meridian (i.e. the same at all longitudes). However, there may be a possibility to use it for close-to edge-on configurations if the precision is sufficient (amplitude greater than 0.3 m/s for inclinations higher than 70$^\circ$, representing about one-third of the RV amplitude, although one should keep in mind that it will be superposed to other contributions to BIS variations  when the star is active.

\subsubsection{Comparison with Earth-like planetary signals}

\begin{figure} 
\includegraphics{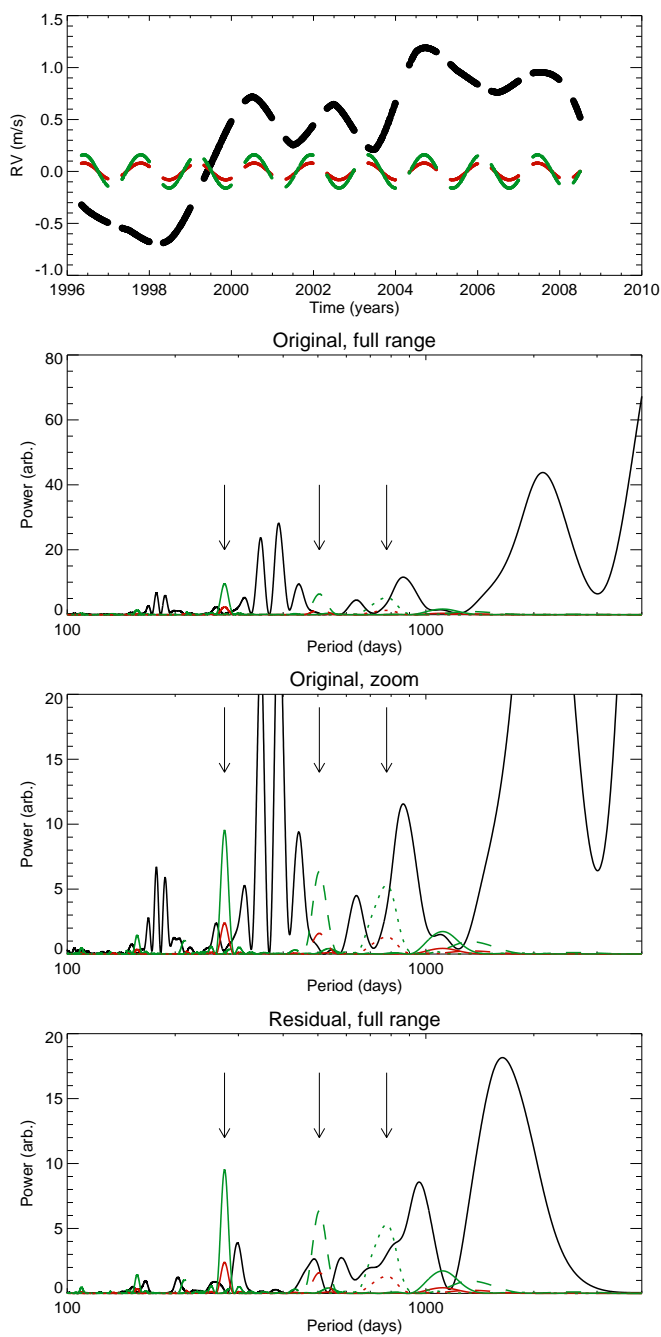}
\caption{
Radial velocity due to edge-on meridional flows vs. time (stars and solid line) and due to 1 (red) and 2 (green) Earth-mass planets in the middle of the habitable zone (first panel). The two following panels show the periodograms for the meridional flow  (black; full range, then focusing on small powers). The planet periodograms  are shown for different orbital periods: inner side of the habitable zone (solid), middle (dashed), and external side (dotted), and the planet peaks are indicated by arrows. 
The last panel is similar but for the residuals after the sinusoidal correction (see text). 
}
\label{periodo}
\end{figure}

We now compare the RV signal due to meridional flows mc(t) with the signal due to an Earth-mass planet in the habitable zone. We consider orbital periods of 274 days (inner side), 505 days (middle of the habitable zone), and 777 days (external side) as in \cite{meunier19b}. For that purpose, we first interpolate the mc(t) time series over a new sampling (the interpolation is performed using a spline function) consisting of 1000 points and covering 12.5 years between 1996 and 2009 (i.e. cycle 23), with a gap of 4 months every year to simulate the fact that a star can in general not be observed at all times. 

Figure~\ref{periodo} (upper panel) shows the time series derived from U10 (edge-on configuration)  and for planets with masses of 1$_{\rm Mearth}$ and 2$_{\rm Mearth}$ in the middle of the habitable zone: the planetary signals are much smaller than the meridional flow signal. The periodograms of the different types of signal are then compared for the two masses and the three orbital periods in the habitable zone, which also shows that such planetary signals are much smaller than the MC signal in this range of periods.

A correction of the stellar signal using a direct linear relationship between RV and the S-index is not useful here because of the strong time lag between the two, as shown in Sect. 2.3.1. A correction using a smoothed version of the S-index for cycle 23 (the smoothing is performed using a simple moving  average) and taking the lag into account for the 14 points of cycle 23 lead to a rms of 0.58 m/s for the residuals  instead of 0.71 m/s, which is a small reduction. Here we use a simple sinusoidal fit, which decreases the rms for the same time series to 0.45 m/s. On the interpolated series shown in Fig.~\ref{periodo}, the rms decreases from 0.57 to 0.26 m/s. This sinusoidal fit of mc(t) significantly decreases the power due to meridional flows, as shown in the lower panel, but the residuals still exhibit significant power for periods in the habitable zone. The exact amplitude of the residuals should however depend on the exact variability on timescales shorter than the cycle period (10s and 100s of days), as discussed in Sect.~2.3.1. The other time series (Appendix A.2 and A.3), although providing slightly different periodograms, lead to similar conclusions. Furthermore, when superposed on other strong signals, a simple sinusoidal fit will not be the best model. More sophisticated techniques will have to be implemented to correct for such flows. 

This example shows the impact of the edge-on configuration. We conclude that in this case, even after the correction described above, the detectability of Earth-like planets in the habitable zone will be challenging, although the performance may be better for such planets on the inner side of the habitable zone. However, specific configurations at inclinations close to the reversal will be more suitable. The impact may also depend on the inclination of the orbital plane of the planet with respect to the equatorial plane, as a 2 M$_{\rm Earth}$ with an inclination of 30$^\circ$ for example corresponds to a 1 M$_{\rm Earth}$ planet seen edge-on. More results taking into account this impact are shown in Sect.~3.

\subsubsection{Superposition to the magnetic activity signal}

\begin{figure*} 
\includegraphics{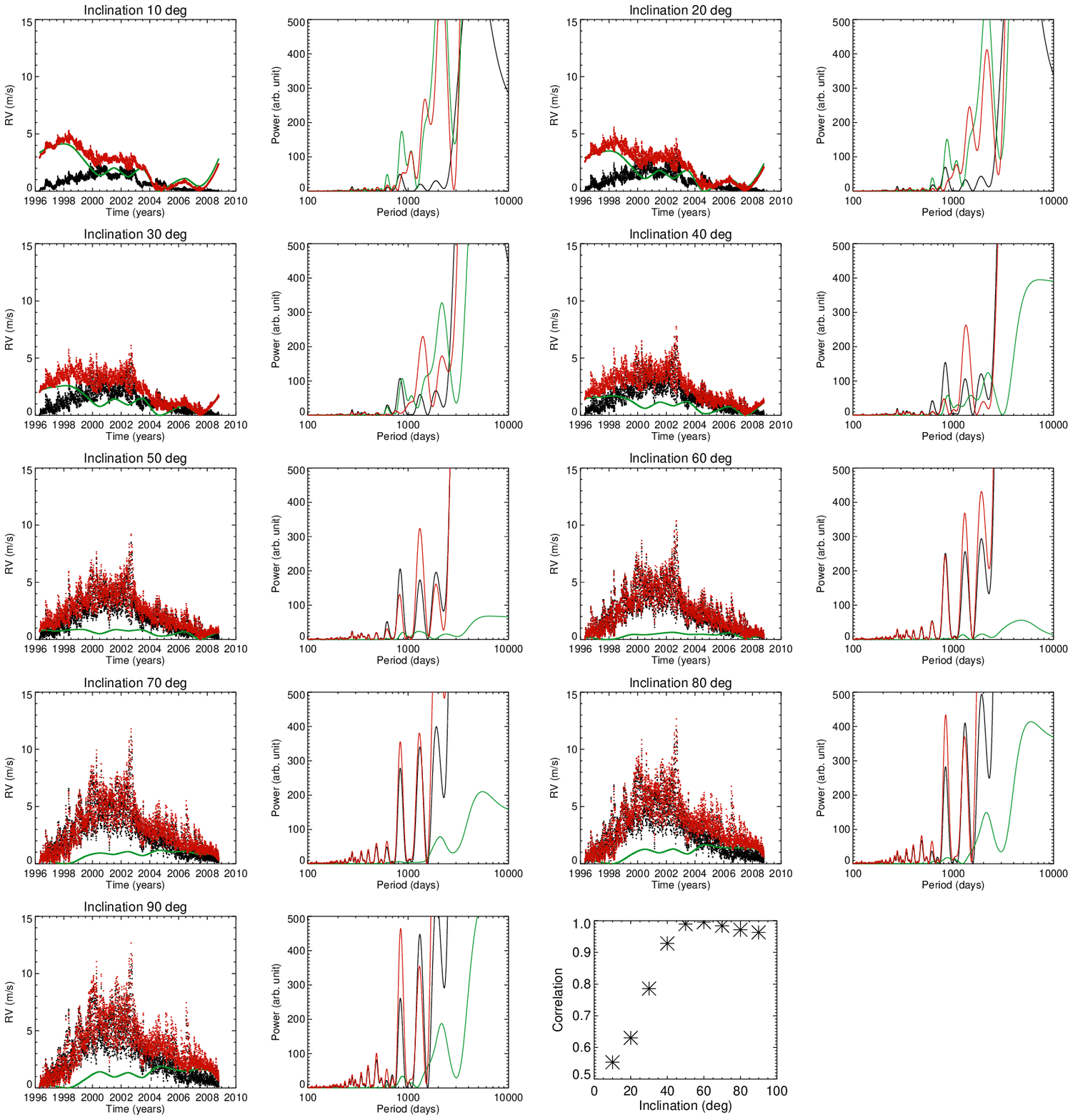}
\caption{
        Radial velocity  due to activity (black dots) and activity superposed on meridional flows (red dots) vs. time for various inclinations, and corresponding periodograms. Meridional flows alone (reconstruction from U10) are shown as a solid green line after interpolation as in Fig.~\ref{periodo}. Offsets are arbitrary (the lowest value for each series has been arbitrarily set to zero). The last plot shows the correlation between the two time series (without and with meridional flows) versus inclination. 
}
\label{superp_incl}
\end{figure*}

\begin{figure} 
\includegraphics{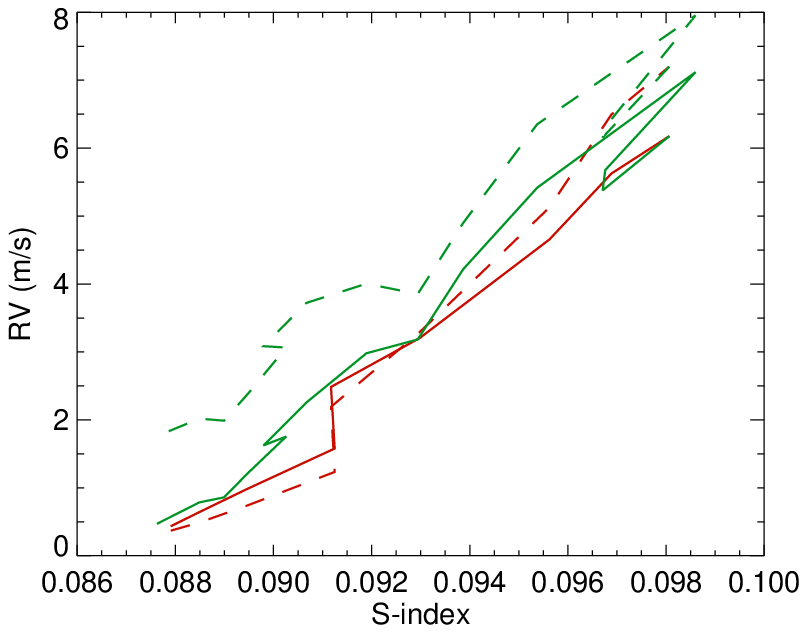}
\caption{
Radial velocity vs. Sacramento Peak Observatory chromospheric emission index, both binned over 6 months, for activity alone (solid lines) and for activity superposed on meridional flows (dashed lines). Red curves correspond to the ascending phase of cycle 23, and green curves to the descending phase. Curves are arbitrarily shifted in RV.
}
\label{superp_hyst}
\end{figure}

Finally, we combine mc(t) to the contribution due to solar magnetic activity, which is dominated by the inhibition of the convective blueshift, but also includes the signal due to spots and plages crossing the disc (through their intensity contrast). We first superpose mc(t) to the magnetic activity RV derived from the solar model of \cite{borgniet15}, which statistically follows cycle 23 general behavior\footnote{The exact magnetic features are different because these RVs result from a model where spots and plages are generated, but they follow cycle 23 amplitude and cycle shape.}: this allows us to consider configurations different from edge-on. Figure~\ref{superp_incl} shows the activity signal alone (in black) and after superposition (in red) with the meridional flow signal for inclinations between 10 and 90$^\circ$. Except for inclinations corresponding to the reversal, the shape is significantly altered, especially for low inclinations, because the meridional flow component, although related to the cycle, is shifted in time and is of strong amplitude. The corresponding periodograms for periods longer than 100 days are also shown. The last plot (lower right panel) of  Fig.~\ref{superp_incl} shows the correlation between the two signals, that is,  before and after superposition of the meridional flow signal, versus inclination. The departure from a correlation of one is low when seen edge-on, while very large departures are present for inclinations below 40$^\circ$. There is no departure from a correlation of one around 55$^\circ$, where the reversal is occurring. 

We now look at the edge-on configuration in more detail. The departure from the original correlation between RV and the S-index is not very large, but it is present. We consider the solar magnetic activity signal reconstructed for cycle 23 in \cite{meunier10a} and compare it to the  Sacramento Peak Observatory S-index (edge-on case only). The correlation between RV and chromospheric emission is computed over 3586 days covering cycle 23, and is equal to 0.91 without the meridional flow (i.e. for magnetic activity alone), and 0.88 after addition of the meridional flow, hence a small decrease. When binning the time series over 6 months, the correlation (sensitive mostly to long timescales) decreases from 0.99 to 0.94. The impact of such departure on the RV signal is illustrated in Fig.~\ref{superp_hyst}, showing the binned RV versus the binned chromospheric emission. As shown in \cite{meunier19c}, even with no MC, this curve presents an hysteresis along the cycle due to projection effects and the butterfly diagram. When taking meridional flows into account (dashed curve), the effect is amplified in this particular configuration (edge-on). The correction method proposed in  \cite{meunier19c}, which was already taking a departure from a linear correlation between RV and chromospheric emission  into account, will therefore have to be adapted and include a time lag introduced by the presence of meridional flows: an additional component of the form $\alpha \times \log R'_{HKsm}$(t+lag), where $\log R'_{HKsm}$ is a smooth version of $\log R'_{HK}$ (to remove the rotational contribution), could be added to their model.

\begin{figure} 
\includegraphics{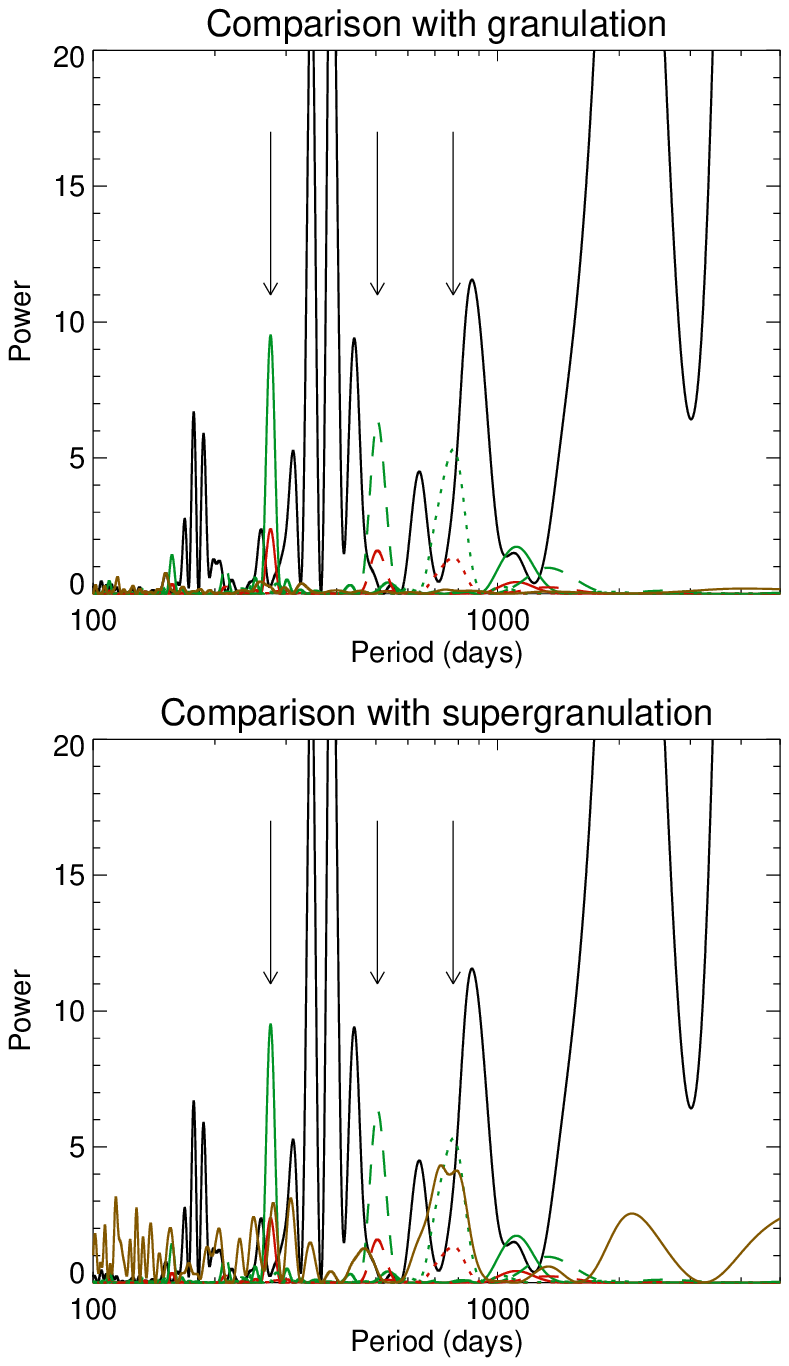}
\caption{
        Periodograms of time series shown in the third panel of Fig.~\ref{periodo} (original time series and planetary RV  for 1 and 2 M$_{\rm Earth}$), added to periodogram due to granulation (upper panel) and supergranulation (lower panel). The colour code for MC and planets is similar to that of Fig.~\ref{periodo}, and the granulation and supergranulation periodograms are shown in brown. 
}
\label{ogs}
\end{figure}

We have so far  compared the MC contribution with the convective blueshift inhibition, which is dominating the long-term variability in the solar case, with a typical amplitude over the cycle of the order of 8 m/s. We now compare the MC amplitude with other sources of stellar signal which have been identified: 
\begin{itemize}
        \item{Spot and plage contrasts: Their contribution mostly affects periods close to the rotation period or its harmonics \cite[e.g.][]{saar97}, with an rms  of the order of 0.33 m/s for the Sun seen edge-on on average over the solar cycle \cite[][]{lagrange10b,meunier10a}. Lower values are obtained at lower inclinations due to projection effects \cite[]{borgniet15}. Their contribution at long periods is therefore much smaller than the MC contribution.}
        \item{Oscillations: The five-minute oscillation signal due to p-modes has a negligible contribution compared to the other stellar signals at long periods and can be easily averaged out \cite[e.g.][]{chaplin19}.  }
        \item{Granulation: The rms of this contribution is estimated to be between 0.4 and 0.8 m/s \cite[][]{elsworth94,palle99,meunier15}. Figure~\ref{ogs} (upper panel) shows a comparison of the periodogram due to granulation after a one-hour binning (the original rms is 0.8 m/s, and 0.39 m/s after binning) for time series similar to those shown in Fig.~\ref{periodo} (MC and planets); the MC contribution is higher than the granulation contribution in the range of periods we are interested in.  }
        \item{Supergranulation: The rms of this contribution is less well determined, but could be between $\sim$0.3 and 1.2~m/s \cite[][]{meunier15}; \cite{palle99} also found a rms of 0.78 m/s.  Figure~\ref{ogs} (lower panel) shows a comparison of the periodogram due to supergranulation (rms of 0.7 m/s, also binned over one hour, which in this case does not significantly change the rms) for time series similar to those shown in Fig.~\ref{periodo} (MC and planets). In this case, although the supergranulation contribution is slightly lower than that from MC, it can lead to peaks of similar amplitudes in certain cases in the range of periods corresponding to the habitable zone.  
                }
\end{itemize}

\section{From F to K stars}

\begin{figure} 
\includegraphics{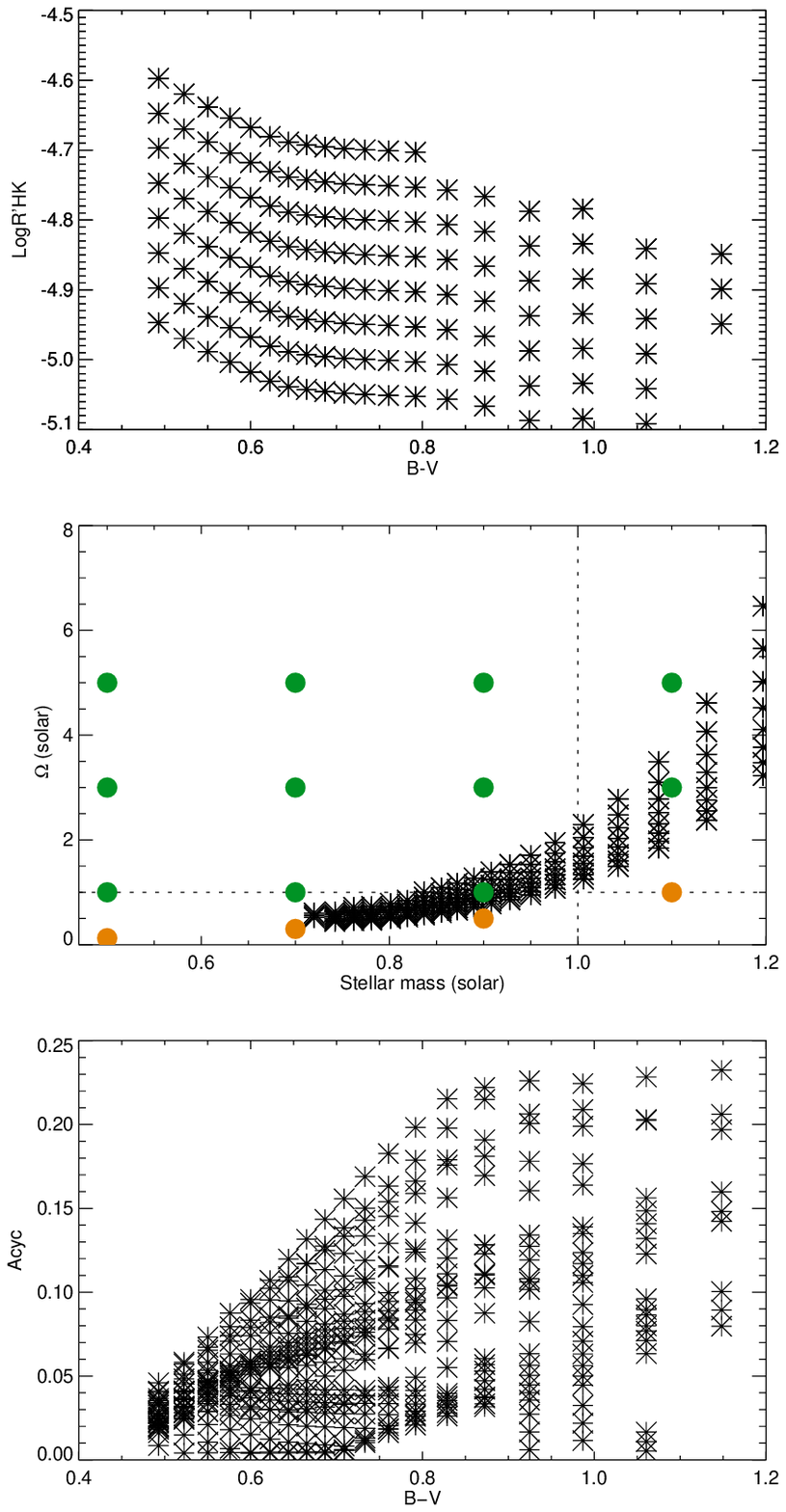}
\caption{
        Average $\log R'_{\rm HK}$ vs. B-V (upper panel), rotation rate $\Omega$ vs. stellar mass (middle panel), and cycle amplitude vs. B-V (lower panel) covered in this work, all extracted from the grid of parameters in \cite{meunier19}. The cycle amplitude is the difference between the prescribed maximum and minimum $\log R'_{\rm HK}$ over the cycle. The green and orange dots in the middle panels represent the position the MHD simulation parameters in \cite{brun17} for solar differential rotation (green) and anti-solar differential rotation (orange). 
}
\label{grille}
\end{figure}

\begin{figure*} 
\includegraphics{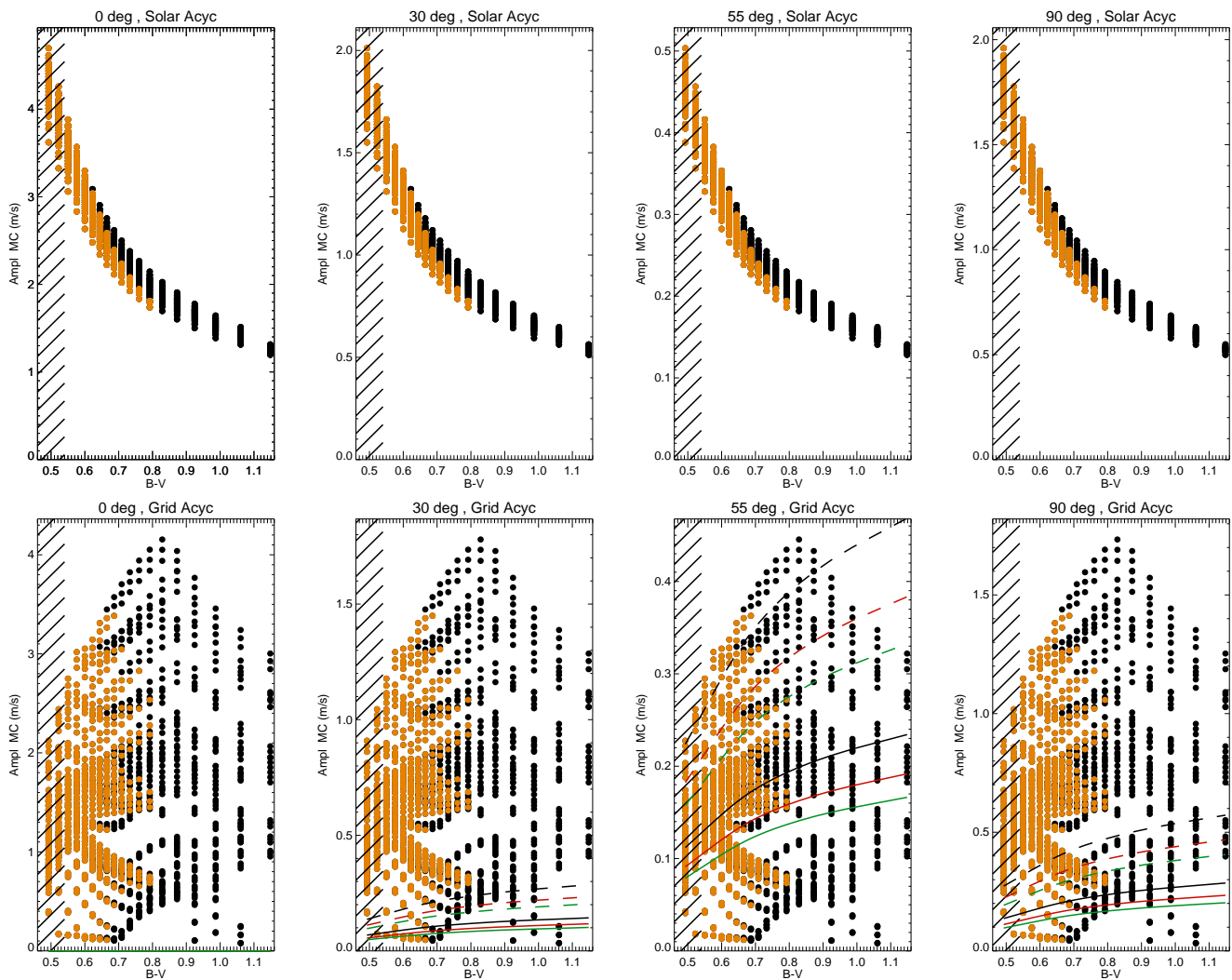}
\caption{
        Stellar RV peak-to-peak amplitude (scaled to solar meridional flows from U10, cycle 23) vs. B-V for solar cycle 23 amplitude (upper panels) and scaled to cycle amplitude from our grid of parameters (lower panels) for four inclinations. Orange symbols represent rotation rates higher than the solar one. The black lines on the left of each panel indicate where the mass is outside the range of hydrodynamical simulation parameters. The peak-to-peak amplitude of the planetary signal, assuming the planet orbital plane is similar to the stellar equatorial plane,  is shown for 1 M$_{\rm Earth}$ (solid lines) and 2 M$_{\rm Earth}$ (dashed lines) and for different orbital periods in the habitable zone: PHZ$_{\rm in}$ (black), PHZ$_{\rm med}$ (red), and  PHZ$_{\rm out}$ (green).    The planetary curves for a pole-on configuration therefore follow the x-axis.
}
\label{stell}
\end{figure*}

\begin{figure} 
\includegraphics{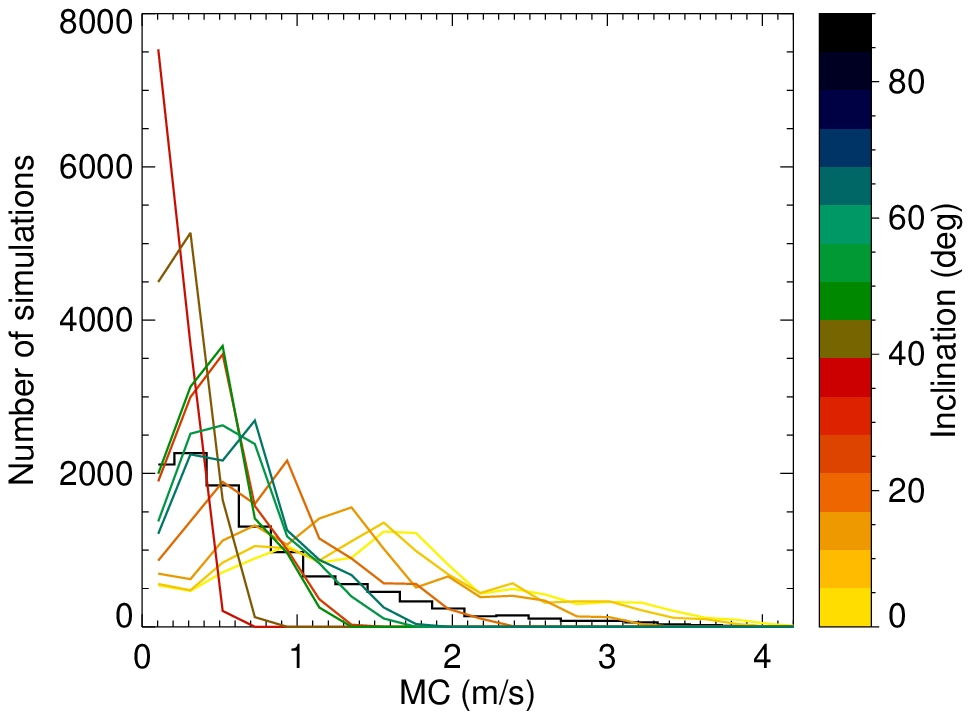}
\caption{
        Distributions of stellar MC values for all simulations (black) and vs. inclinations (from pole-on in yellow to edge-on in blue,  as shown by the colour bar). 
}
\label{stell_dist}
\end{figure}

In this section, we estimate the amplitude of the integrated meridional flows for a range of stars similar to the magnetic activity simulations in \cite{meunier19}, i.e. covering the spectral range F6-K4 and activity levels corresponding to old main sequence stars. The range of parameters is shown in Fig.~\ref{grille}. Stellar RVs are expected to depend on spectral type and rotation rate (i.e. in our case on activity level). Since the variability of the solar integrated MC obtained in Sect.~2 is related to the cycle, stellar MC amplitudes should also depend on the cycle amplitude. To our knowledge, there has so far not been any observation of these flows for stars other than the Sun because they are very weak. However, a number of 3D hydrodynamical simulations have been performed in several studies, which allows us to scale the stellar RV to the solar one. We then study the effect of other properties, such as for example the likely presence of multi-cellular patterns for certain stars, and the impact of the limb darkening function. The MC amplitude is compared to the amplitude of Earth-like planets in the habitable zone and to the RV amplitude due to the inhibition of the convective blueshift (the dominating part due to magnetic activity over the cycle) to quantify the importance of the MC contribution.

\subsection{Scaling of the solar amplitudes with spectral type and Prot}

In this section, we first consider the effect of rotation rate and mass on the global meridional flow amplitude. \cite{ballot07} simulations (made for 1 M$_\odot$ only) showed that the amplitude of the meridional flows decreases for smaller rotation periods (i.e. larger values of $\Omega$). For a given mass, stars are strongly affected by stellar activity, which is proportional to the rotation rate: the quietest stars have the longest rotation periods and therefore we expect them to exhibit the highest meridional circulation. This means that in our grid of parameters, for a given mass, the quietest stars should show the greatest meridional circulation. \cite{ballot07} modelled its amplitude with a scaling law in $\Omega^{\beta}$, where the exponent $\beta$  took values between -0.53 and -1.02 depending on the set of simulations. In addition, \cite{matt11} obtained larger meridional flows for high-mass stars for a solar rotation rate. Both studies correspond to a limited range of parameters, and these works were subsequently extended to a larger range of parameters in \cite{brun17}, who obtained similar trends. The range of parameters covered in \cite{brun17} is shown in Fig.~\ref{grille} (middle panel, orange and green dots). Most of the parameters we consider correspond to solar differential rotation, as discussed in \cite{meunier19}, which is also apparent in this figure (which indicates which hydrodynamical simulations correspond to solar or anti-solar differential rotation). We model the meridional flow amplitude obtained by \cite{brun17} as proportional to $\Omega^{\beta}M^{\alpha}$ from the green dots (i.e. solar differential rotation only). Adjusting the power law in $\Omega$ for each of the four ranges in mass leads to $\beta$ varying between -0.19 and -0.36, with an average of -0.26. The dependence on the mass can then be modelled with $\alpha$=3.44. The results are not very different when considering all simulations (both solar and anti-solar differential rotation).

We apply these scaling laws (with $\alpha$=3.44, $\beta$=-0.26) to the solar peak-to-peak amplitudes (the values corresponding to rms RV are smaller by about a factor two; see Table~\ref{tab_rms}) of mc(t) derived in Sect.~2 for the U10 solar profiles (cycle 23) and four inclinations: 0$^\circ$ (pole-on), 30$^\circ$, 55$^\circ$ (where the variability is the lowest), and 90$^\circ$ (edge-on). The results corresponding to a solar cycle amplitude are shown in the upper panels of Fig.~\ref{stell}. The dependence is dominated by the mass, while the effect of the rotation is weaker, within the range of parameters we consider. 

We then scale the MC amplitude with the amplitude of the cycle in our grid of simulations as described in \cite{meunier19}, following the prescription shown in Fig.~\ref{grille} (lower panel) and assuming the amplitude of the meridional flow variability is linearly correlated with the cycle amplitude. These results are presented in the lower panels of Fig.~\ref{stell}, again for four inclinations.
In this case, the meridional flow amplitudes are dominated by the cycle amplitude, that is, by the magnetic activity variability. Fine details in the scaling laws describing how the meridional flows vary with mass and rotation rate are therefore not as critical here. More work on how the meridional flows depend on the magnetic activity variability will be needed in the future however, since we have considered a linear relationship here. 

As a consequence, there is no strong trend of the expected MC amplitude versus mass once it is scaled by the activity amplitude, which shows a very large dispersion. This is because each given spectral type has a large range of possible activity levels. 
For the edge-on configuration, the expected peak-to-peak amplitude takes values up to 1.8 m/s (edge-on), with many values in the 0.2-1 m/s range. 
For the pole-on configuration, the amplitude can be as high as 4 m/s. Considering stellar inclination where the integrated RV is reversing would lead to much lower amplitudes (lower than 0.5 m/s). 
The distributions of peak-to-peak amplitudes for different inclinations is shown in Fig.~\ref{stell_dist}. The maximum of the global distribution is around 0.3 m/s. For edge-on configurations, the maximum of the distribution is in the range 0.3-0.8~m/s.

In this section, we directly scale the solar flow (obtained with a limb darkening function corresponding to the solar T$_{\rm eff}$) using the values obtained in Sect.~2. We check what would be the impact of a limb-darkening function computed for the proper temperature over the range of parameters on the integrated meridional flow.  Trends are observed with stellar type as expected, but we find that the effect of the limb-darkening function can be neglected, with a maximum departure of the order of 0.8\% (for the edge-on configuration, and B-V around 1.05). We therefore neglect this effect here. 

 For fast rotating stars (typically faster than the Sun), hydrodynamical simulations show that the latitudinal profiles could be more complex than the solar one, and exhibit a multi-cellular pattern \cite[][]{matt11,guerrero13,guerrero16}. We investigated the impact of such a pattern and found that this tends to reduce the RV effects induced by MC due to cancellation between different latitudes. However, such fast rotating stars are  not very good targets to search for low-mass planets due to their faster rotation than the Sun, because of the lower achievable RV precision.

\subsection{Comparison with Earth-like planetary signals}

\begin{figure} 
\includegraphics{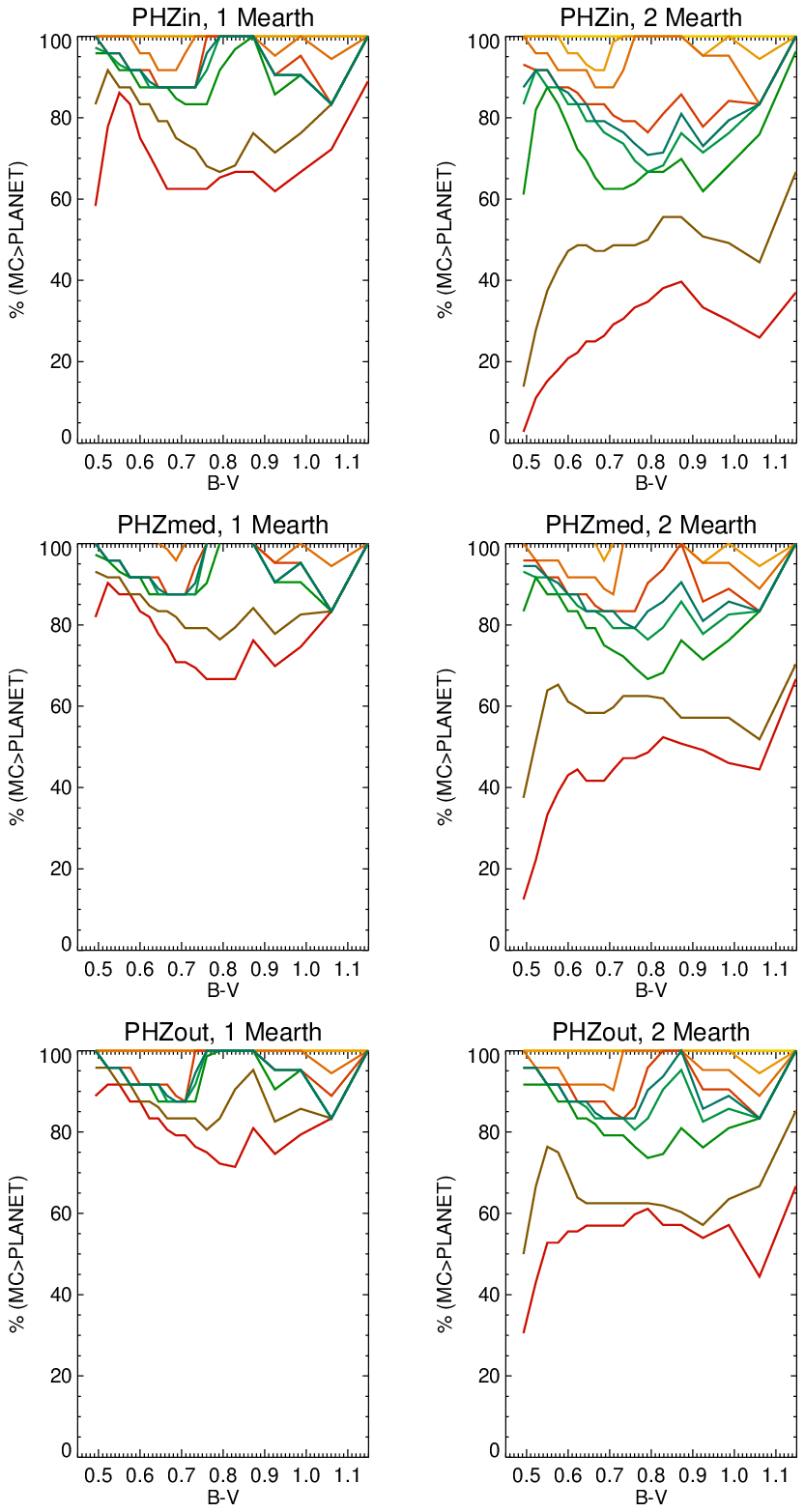}
\caption{
Percentage of simulations for which MC is higher than planetary signal vs. B-V. Each panel corresponds to a different mass and orbital period. The colour code is similar to Fig.~\ref{stell_dist}: inclination from pole-on in yellow to edge-on in blue (see colour bar on Fig.~\ref{stell_dist}). Red corresponds to 50$^\circ$.  
}
\label{pcpla}
\end{figure}

\begin{table}
        \caption{Percentage of simulations with MC larger than planetary signal}
\label{tab_pla}
\begin{center}
\renewcommand{\footnoterule}{}  
\begin{tabular}{c|c|c|c|c|c|c}
\hline
        PHZ  &  \multicolumn{3}{c|}{1 M$_{\rm Earth}$} & \multicolumn{3}{c}{2 M$_{\rm Earth}$} \\
        \hline
  &  0$^\circ$ & 55$^\circ$ & 90$^\circ$ & 0$^\circ$ & 55$^\circ$ & 90$^\circ$ \\
\hline
        PHZ$_{\rm in}$ &  100 & 45.9 & 87.7 & 100 & 3.2 & 69.5 \\
        PHZ$_{\rm med}$ &  100 & 61.2 & 90.9 & 100 & 10.2 & 76.1 \\
        PHZ$_{\rm out}$ &  100 & 67.0 & 92.4 & 100 & 19.1 & 80.1 \\
\hline
\end{tabular}
\end{center}
\end{table}

Figure~\ref{stell} also shows the peak-to-peak amplitude of planetary signals for 1 M$_{\rm Earth}$ and 2 M$_{\rm Earth}$ planets orbiting in the habitable zone (three values of the orbital periods). This amplitude can be compared with the MC contribution without noise. The amplitude shown here is projected according to the assumption that the orbital plane is similar to the stellar equatorial plane. As a consequence, the amplitude is zero for pole-on configurations (which are not very suitable for detecting planets). However, there could be departures from these curves if there is an inclination between the two planes: in this case, the maximum planetary amplitude is the one shown in the panel corresponding to the edge-on configuration.  
In most cases, the MC amplitude is larger than an Earth-mass planet in the habitable zone, except for a small number of simulations corresponding to very quiet stars seen with an intermediate inclination.
The percentage of simulations above those planetary levels is shown in Table~\ref{tab_pla} for three values of habitable zone periods (PHZ) and two planet masses, again assuming  that the orbital plane and the equatorial plane are similar. 
These percentages are lowest only for 2 M$_{\rm Earth}$ and intermediate inclinations. 
All percentages are shown in Fig.~\ref{pcpla}; for 1 M$_{\rm Earth}$, they are never lower than 70\% and are their lowest for inclinations around 50$^\circ$ (red curves) and 60$^\circ$ (brown curves). Percentages are lower for 2 M$_{\rm Earth}$ and can be close to zero for these intermediate inclinations, although the percentage is always above 60\% for other configurations.

\subsection{Comparison to the RV amplitude due to magnetic activity}

\begin{figure} 
\includegraphics{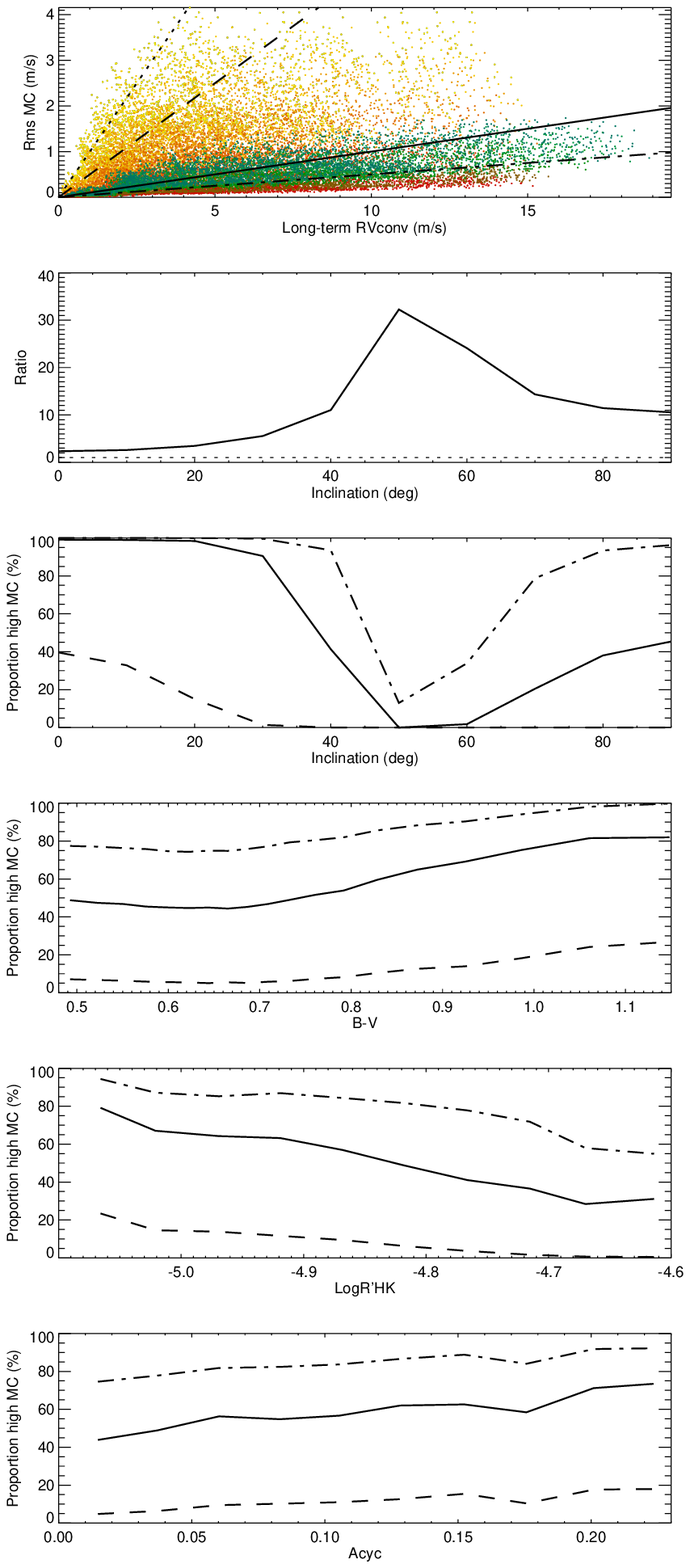}
\caption{
        Meridional circulation amplitude vs. long-term amplitude of convective blueshift inhibition (upper panel) on our grid of parameters \cite[from][]{meunier19} for different inclinations (from pole-on in yellow to edge-on in blue). Only one point in five is shown for clarity. The straight lines correspond to MC amplitude representing 5\% (dotted-dashed line), 10\% (solid line), 50\% (dashed line), and 100\% (dotted line) of the convective blueshift inhibition. The second panel shows the median ratio between the convective blueshift inhibition component and MC vs. inclination. The four following panels show the percentage of simulations in which  MC represents more than 5\% (resp. 10\%, 50\%) (in dotted-dashed lines, resp. solid lines, dashed lines) of the convective blueshift amplitude vs. inclination, B-V, $\log R'_{\rm HK}$, and A$_{\rm cyc}$.  
}
\label{comp_conv}
\end{figure}

\begin{figure} 
\includegraphics{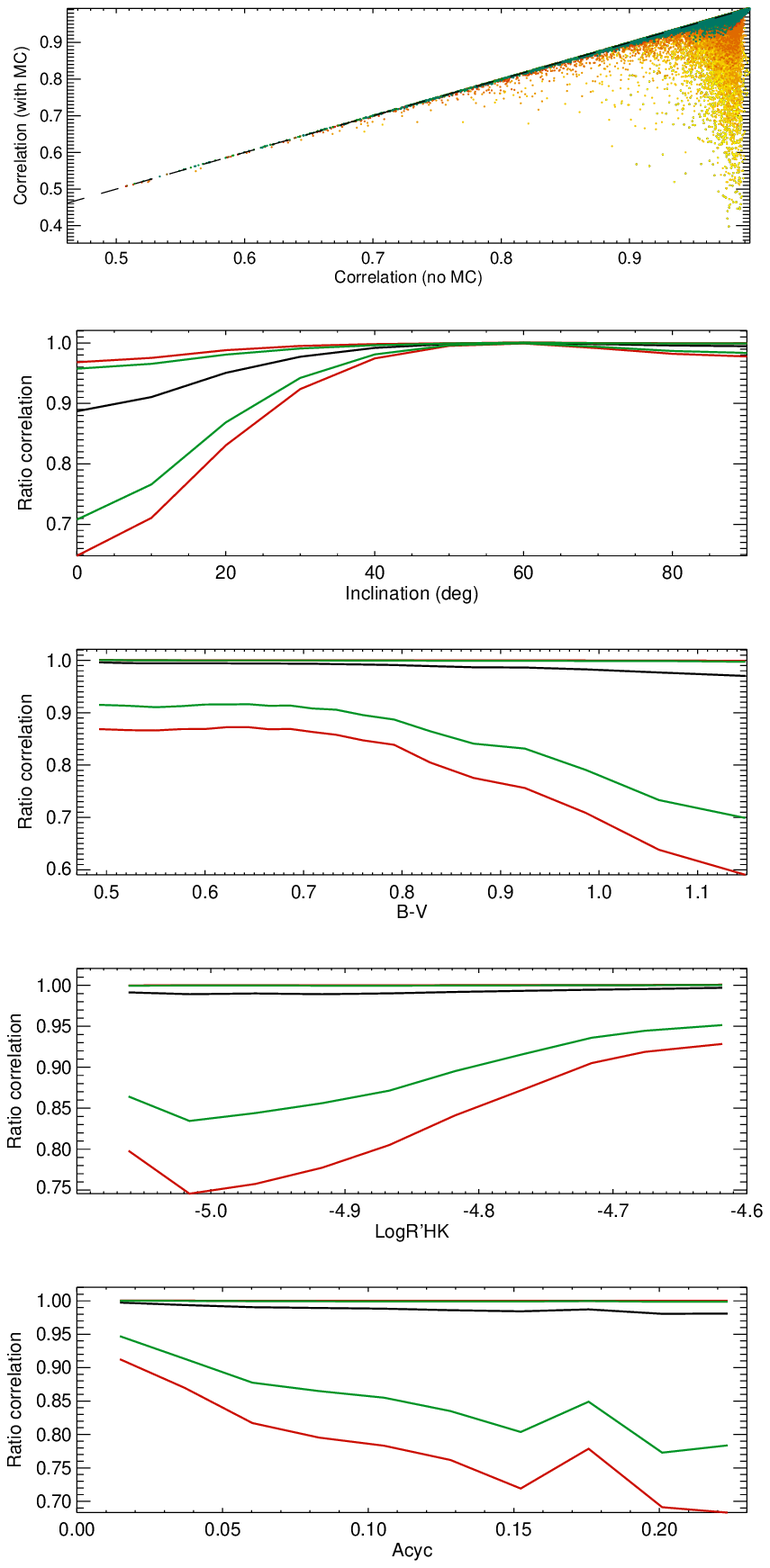}
\caption{
        Correlation between activity+MC signal and $\log R'_{\rm HK}$ vs. correlation with activity alone. The colour code represents inclination as in Fig.~\ref{comp_conv}.  
        The four following panels show the median ratio between correlation for activity+MC divided by correlation for activity alone vs. inclination, B-V, $\log R'_{\rm HK}$, and A$_{\rm cyc}$ (red lines). The green lines represent the 5th and 95th percentiles, and the dotted lines the 10th and 90th percentiles.
}
\label{comp_correl}
\end{figure}

A detailed analysis of the effect of MC on exoplanet detectability is beyond the scope of this paper because it requires the inclusion of magnetic activity in order to be realistic given the distortion of the variability introduced by this latter (see Sect. 2.3.4). However, it is interesting to estimate how the amplitudes we have found compare with the other processes leading to significant long-term variability, such as the inhibition of the convective blueshift obtained in \cite{meunier19}, which we know is very challenging for exoplanet detections. The comparison between peak-to-peak MC amplitudes and the long-term amplitude of the convective blueshift inhibition obtained in \cite{meunier19} is shown in Fig.~\ref{comp_conv} for all inclinations (upper panel). We find that in general the convective blueshift contribution is larger, except for a few configurations for inclinations lower than 30$^\circ$. 
Only 1.3\% of the simulations correspond to MC amplitude larger than the convective blueshift inhibition amplitude. 
We compute the proportion of simulations for which the MC amplitude represents at least a certain fraction (5\%, 10\%, and 50\%) of the convective blueshift amplitude (the straight lines in the first panel indicate those proportions). 
These proportions are 8.9\%, 53.4\%, and 80.8\% for the 50\%, 10\%, and 5\% levels.  
The median ratio over all realisations  is 9.2, with ratios for individual realisations between 0.5 and 208.  
 We also computed the median ratio separately for each inclination, as shown in the second panel of Fig.~\ref{comp_conv}: the median ratio at a given inclination is always higher than 1 (although not very far from it for low inclinations), and is significantly higher for inclinations around 40-50$^\circ$ (median ratio higher than 30): those inclinations are the most suitable configurations for exoplanet detectability, from the point of view of meridional flows. 

The percentage of simulations where MC represents more than a certain fraction of the convective blueshift inhibition is shown in the following panels versus different variables: inclination, B-V, $\log R'_{\rm HK}$, and A$_{\rm cyc}$.  
The 50\% level is usually low, with a predominance of low inclinations, and to a lesser extent of quiet, low-mass stars: the percentage of simulations where MC represents more than half of the convective blueshift inhibition can be as high as 40\% for an inclination of 0$^\circ$, showing again the possibility for strong distortions of the RV--chromospheric emission relationship. A large proportion of simulations have MC representing more than 10\% of the convective blueshift inhibition amplitude (except around 50-60$^\circ$), which should lead to significant distortions of the relationship between RV and activity indicators: this will therefore affect corrections and exoplanet detectability. The most suitable configurations are again medium inclinations, followed by high mass, more active stars (with $\log 
R'_{HK}$ above $\sim$-4.7), at least from the point of view of the relative contribution of meridional flows compared to the inhibition of the convective blueshift.

Finally, we estimate the distortion of the RV-$\log R'_{\rm HK}$ relationship brought about by the presence of MC. For each simulation, we estimate the meridional circulation by smoothing the $\log R'_{\rm HK}$ time series, shifting it with a time lag similar to the solar one (0.23 times the cycle period), and scaling its amplitude as in Sect.~3.1. The MC signal is then added to the magnetic activity contribution of \cite{meunier19} and the correlation between RV (before adding MC and after) and $\log R'_{\rm HK}$ is computed.  The results are shown in Fig.~\ref{comp_correl}.
As for the previous comparisons, the main effects are seen for low inclinations (the median ratio,  where the ratio is between the correlation with activity+MC and the correlation with activity alone, is down to 0.9 for pole-on stars), especially for quiet (on average), low-mass   stars, mostly when exhibiting some variability.

\section{Conclusion}

We reconstructed the integrated solar RV due to meridional flows with very good temporal coverage (two cycles, all inclinations). The MC variability is clearly related to the cycle, although cycles of similar amplitude (cycles 22 and 23) lead to different amplitudes: the difference is of the order of 50\%, which is most likely the main source of difficulty that we face when attempting to predict expected amplitudes. The most
interesting time series  for exoplanet detectability is the reconstruction made from latitudinal profiles obtained from Dopplergram analysis \cite[][]{ulrich10} because they correspond to flows similar to those seen when measuring RV in the visible. The integrated RV exhibits a time lag with respect to the cycle. The other reconstructions (using results obtained with ring diagram helioseismology, and magnetic feature tracking) are more closely anti-correlated with the cycle (maximum at activity minimum and vice versa), and amplitudes are robust.
The integrated RV also varies on shorter timescales, but the exact behaviour of the integrated RV due to meridional flows at these scales is still unclear. 

Typical values for the rms of the signal (over cycle 23) are of the order of 0.7 m/s for the edge-on configuration and 1.6 m/s for the pole-on configuration (cycle 23), with a minimum around 0.3 m/s for an inclination of 55$^\circ$ (where there is a reversal). Peak-to-peak amplitudes are respectively of 1.4 and 3.3 m/s for the edge-on and pole-on configurations. 
The exact position of the reversal may vary depending on the exact shape of the latitudinal profile over time (45$^\circ$ for cycle 22).  

We compared these solar time series with the signal that would be due to an Earth-like planet in the habitable zone around the Sun: the power due to meridional flows is much larger than the planetary signal. Even after a sinusoidal fit to correct for the meridional flow contribution, the power at periods corresponding to the habitable zone is still large  and above the level corresponding to Earth-like planets. Furthermore, when added to the magnetic activity signal, the correlation existing between RV due to magnetic structures (in particular the inhibition of the convective blueshift) and chromospheric emission is degraded. The degradation is moderate for edge-on configurations (but must be taken into account to reach very low detection limits). We find that the hysteresis pattern observed between these two observables \cite[][]{meunier19c} is amplified by the presence of meridional circulation due to the time lag (around 2.5 years) between RV due to meridional flows and the cycle. The effect is naturally very small when reaching the reversal around 55$^\circ$, but becomes major at lower inclinations, leading to RV variations that show departure from a good correlation with the chromospheric emission variations.  This alteration of the RV-chromospheric emission relationship should be a major difficulty to correct for this signal as well as to correct for magnetic activity in such conditions, especially for configurations close to pole-on. Nevertheless, we note that the pole-on configuration is not the most suitable for exoplanet detection in RV, because the signal of the planet falls to zero if the orbital plane is similar to the stellar equatorial plane.  The bisector variations may be significant (up to 0.4-0.45 m/s) for configurations close to edge-on, which could be used to remove part of the MC contribution in those cases.

Finally, using scaling laws from the numerical simulations of \cite{brun17}, we extrapolated the amplitude of the integrated meridional flows that we find for the Sun to other solar-type stars. For an  amplitude of the activity cycle similar to the solar one (i.e. considering only a scaling in mass and rotation), the variability is dominated by the mass, with a much larger RV variability for F stars compared to K stars. However, if we also scale this variability with the amplitude of the cycle expected for different types of stars while assuming a linear relationship valid for all stars then the variability is in fact dominated by the cycle amplitude. In view of  the differences of up to 50\% that we see for the Sun, it will be very important to better understand this relationship in the future. For stars seen edge-on, the peak-to-peak amplitude can reach 1.8 m/s, and up to 4 m/s for pole-on stars. At stellar inclinations close to the reversal, the amplitudes are much smaller. 
Furthermore, some stars may exhibit multi-cellular patterns, as shown from numerical simulations: this should significantly decrease the RV amplitude, and so these configurations may be significantly more suitable. We expect this effect to play a role for F stars and early G stars, but not for late G and K stars, in the range of rotation periods we consider here.

The present study focuses on characterising the amplitude of the MC contribution. We find that its variability is of the same order of magnitude as or slightly larger than other processes (granulation, supergranulation, spots, plages), i.e. typically in the range 0.5-1 m/s for a star with a solar-like cycle, and can be much higher for  stars with low inclination. It also represents a significant proportion of the amplitude of the signal due to the inhibition of the convective blueshift, which will have a significant impact on exoplanet detectability, as shown in Sects.~2.3.5 and 3.4. 
However, we have not estimated the performance in terms of detection rates and false-positive levels because this  must be done in conjunction with the presence of magnetic activity
in order to obtain meaningful results, which is beyond the scope of this paper: this will be done in a future study, following a systematic approach combining MC and magnetic activity contributions using blind tests.

\begin{acknowledgements}

We thank  S. Brun and A. Strugarek for usefull discussions about this work. 
This work has been funded by the ANR GIPSE ANR-14-CE33-0018.
This work was supported by the "Programme National de Physique Stellaire" (PNPS) of CNRS/INSU co-funded by CEA and CNES.
This work was supported by the Programme National de Plan\'etologie (PNP) of CNRS/INSU, co-funded by CNES.
The monthly sunspot number has been provided by 
the SIDC-team, World Data Center for the Sunspot Index, Royal Observatory of Belgium (http://www.sidc.be/sunspot-data/). 
This work made use of the Ca K index  provided by the Sacramento Peak Observatory of the U.S. Air Force Phillips Laboratory.

\end{acknowledgements}

\bibliographystyle{aa}
\bibliography{biblio}

\begin{appendix}

\section{Integrated meridional-flow time series}

\subsection{Table with the reconstructed RV from \cite{ulrich10}}

Table~\ref{tab_u10} provides the integrated RV due to MC that we reconstructed from the latitudinal profiles of \cite{ulrich10}. Lower and upper estimates of uncertainties are also provided. The results are discussed in Sect.~2.3.1.

\begin{table*}
\caption{Integrated meridional flow from \cite{ulrich10}}
\label{tab_u10}
\begin{center}
\renewcommand{\footnoterule}{}  
\begin{tabular}{llllllllll}
        \hline
        Time   & RV at 0$^\circ$  & $\sigma_{\rm high}$ at 0$^\circ$ & $\sigma_{\rm low}$ at 0$^\circ$& RV at 55$^\circ$ & $\sigma_{\rm high}$ at 55$^\circ$ & $\sigma_{\rm low}$ at 55$^\circ$  & RV at 90$^\circ$   & $\sigma_{\rm high}$ at 90$^\circ$& $\sigma_{\rm low}$ at 90$^\circ$  \\
        (year) & (m/s)   & (m/s) & (m/s)& (m/s)& (m/s)& (m/s)& (m/s)       & (m/s) & (m/s)   \\    \hline
1986 &  -6.45 &   0.51 &   0.05 &   0.12 &   0.02 &   0.05 &   3.88 &   0.25 &   0.02 \\
1987 &  -6.05 &   0.49 &   0.06 &   0.07 &   0.02 &   0.05 &   3.51 &   0.26 &   0.03 \\
1988 &  -5.89 &   0.48 &   0.05 &  -0.52 &   0.01 &   0.04 &   3.38 &   0.25 &   0.02 \\
1989 &  -6.38 &   0.56 &   0.07 &  -0.20 &   0.02 &   0.06 &   3.72 &   0.30 &   0.02 \\
1990 &  -5.75 &   0.69 &   0.08 &  -0.09 &   0.04 &   0.06 &   3.39 &   0.35 &   0.03 \\
1991 &  -8.09 &   0.58 &   0.07 &   0.32 &   0.04 &   0.05 &   4.50 &   0.31 &   0.02 \\
1992 &  -8.51 &   0.53 &   0.05 &   0.50 &   0.04 &   0.05 &   4.70 &   0.27 &   0.02 \\
1993 &  -7.66 &   0.52 &   0.07 &   0.32 &   0.01 &   0.04 &   4.22 &   0.28 &   0.02 \\
1994 &  -7.94 &   0.49 &   0.05 &  -0.06 &   0.01 &   0.03 &   4.29 &   0.25 &   0.01 \\
1995 &  -8.09 &   0.50 &   0.04 &  -0.33 &   0.01 &   0.03 &   4.34 &   0.26 &   0.02 \\
1996 &  -7.24 &   0.69 &   0.09 &  -0.06 &   0.01 &   0.05 &   3.96 &   0.32 &   0.02 \\
1997 &  -6.67 &   0.51 &   0.06 &  -0.19 &   0.01 &   0.04 &   3.78 &   0.25 &   0.02 \\
1998 &  -6.78 &   0.59 &   0.06 &  -0.10 &   0.02 &   0.04 &   3.69 &   0.31 &   0.03 \\
1999 &  -8.24 &   0.54 &   0.06 &   0.08 &   0.02 &   0.04 &   4.41 &   0.27 &   0.03 \\
2000 &  -9.47 &   0.52 &   0.07 &  -0.09 &   0.01 &   0.05 &   5.06 &   0.27 &   0.02 \\
2001 &  -8.84 &   0.53 &   0.08 &   0.15 &   0.02 &   0.05 &   4.59 &   0.28 &   0.02 \\
2002 &  -9.64 &   0.52 &   0.04 &   0.15 &   0.01 &   0.04 &   4.99 &   0.27 &   0.02 \\
2003 &  -8.87 &   0.51 &   0.06 &   0.10 &   0.01 &   0.04 &   4.55 &   0.26 &   0.02 \\
2004 & -10.87 &   0.49 &   0.05 &  -0.15 &   0.01 &   0.04 &   5.49 &   0.25 &   0.02 \\
2005 & -10.31 &   0.54 &   0.06 &  -0.15 &   0.05 &   0.05 &   5.31 &   0.27 &   0.02 \\
2006 &  -9.81 &   0.45 &   0.04 &   0.08 &   0.03 &   0.03 &   5.10 &   0.23 &   0.02 \\
2007 & -10.45 &   0.44 &   0.05 &  -0.52 &   0.07 &   0.03 &   5.29 &   0.14 &   0.02 \\
2008 &  -8.84 &   0.50 &   0.06 &  -0.17 &   0.02 &   0.05 &   4.86 &   0.25 &   0.02 \\
2009 &  -5.57 &   0.51 &   0.04 &  -0.51 &   0.00 &   0.04 &   3.13 &   0.26 &   0.02 \\
\hline
\end{tabular}
\end{center}
        \tablefoot{Low and high estimates of uncertainties correspond to two assumptions, see text in Sect.~2.3.1. 
}
\end{table*}

\subsection{MC from \cite{basu10}}

\cite{basu10} (hereafter BA10) performed a ring diagram analysis of Dopplergrams from MDI/SOHO data from between 1996 and 2009. This technique probes flows in different layers below the photosphere down to about 10 Mm. Here we use the flows corresponding to the  closest layer to the surface, i.e. 2.8 Mm deep and the OLA (Optimally Localized Average) inversion (two inversions are provided in BA10, but they give very similar results). BA10 decomposed each latitudinal profile into a sum of associated Legendre polynomials: we digitised these coefficients versus time to reconstruct the profiles. Four even coefficients (corresponding to the antisymmetric component of the flows, c2 to c8) were used for a good reconstruction, and are on average equal to 9.5, -3.6, -0.8, and  -2.2 m/s respectively, with uncertainties on average of the order of 0.5, 0.7, 0.7, and 0.5 m/s respectively. The resulting time series mc(t) are given in Table~\ref{tab_ba10} for 0$^\circ$, 55$^\circ$, and 90$^\circ$.

\begin{table}
\caption{Integrated meridional flow from \cite{basu10}}
\label{tab_ba10}
\begin{center}
\renewcommand{\footnoterule}{}  
\begin{tabular}{llll}
        \hline
Time   & RV at 0$^\circ$  & RV at 55$^\circ$   & RV at 90$^\circ$   \\
(year) & (m/s)     & (m/s) & (m/s)   \\  \hline
1996 &  -9.87 &   0.01 &   5.10 \\
1997 &  -7.08 &  -0.10 &   3.91 \\
1998 &  -8.15 &  -0.06 &   4.45 \\
1999 &  -5.22 &  -0.26 &   3.31 \\
2000 &  -4.11 &  -0.30 &   2.86 \\
2001 &  -5.82 &  -0.08 &   3.26 \\
2002 &  -4.99 &  -0.17 &   2.94 \\
2002 &  -3.34 &  -0.35 &   2.51 \\
2003 &  -6.61 &  -0.19 &   3.84 \\
2003 &  -6.49 &  -0.09 &   3.56 \\
2004 &  -7.61 &  -0.17 &   4.27 \\
2005 &  -7.74 &  -0.19 &   4.39 \\
2006 &  -8.47 &  -0.03 &   4.44 \\
2008 &  -9.25 &   0.03 &   4.72 \\
2009 &  -9.41 &  -0.03 &   4.97 \\
\hline
\end{tabular}
\end{center}
\end{table}

\subsection{MC from \cite{hathaway10} and \cite{hathaway11}}

\cite{hathaway10} (hereafter HR10) used a very different method: they estimated the meridional flows by tracking magnetic features in magnetograms from MDI/SOHO from between 1996 and 2009 because they were interested to see the flows actually seen by these magnetic features. This was crucial to make better estimates of the contribution of these flows in flux transport dynamo models, in which they play a very significant role. We use their time average latitudinal profile, which is then scaled by the amplitude versus time, both digitised from HR10. Unlike the results from U10 and BA10, which were averaged over one year, the temporal cadence is much better here: the MC latitudinal profiles  are averaged over the rotation period instead of one year. The use of the averaged profile means that any converging flow pattern superposed on the relatively smooth profile is not taken into account here. These flows, in addition to being weaker than those derived by U10, also present less polar reversals. This may be due to the fact that they may probe different layers (below the photosphere) and/or may represent how the magnetic flux tubes move across the flows, as discussed by \cite{ulrich10}. 

\cite{hathaway11} (hereafter HR11) use the same technique, and achieve results that are only slightly different and with a slightly longer temporal coverage (1996-2010). These authors modelled the latitudinal profiles with two functions in latitude: we digitised the two corresponding coefficients S$_1$(t) and S$_3$(t) and reconstructed the profiles using:
\begin{equation}
mc(t,\theta) = S_1(t) (2 \sin(\theta)\cos(\theta)) + S_3(t) (7\sin^2(\theta)- 3\sin(\theta))\cos(\theta). 
\end{equation}
As in HR10, the description of the flows does not include complex patterns such as the converging flows in the activity belt. The coefficients S$_1$(t) and S$_3$(t) are on average equal to 11 m/s and -2.2 m/s respectively (with dispersion over time of the order of 2 m/s), and the typical uncertainties are 0.9 m/s and 0.4 m/s respectively. 
We  only show the table (Table~\ref{tab_hr11}) for \cite{hathaway11} since the results are very similar to those from \cite{hathaway10}, but slightly longer.

\subsection{Comparison between time series}

\begin{figure} 
\includegraphics{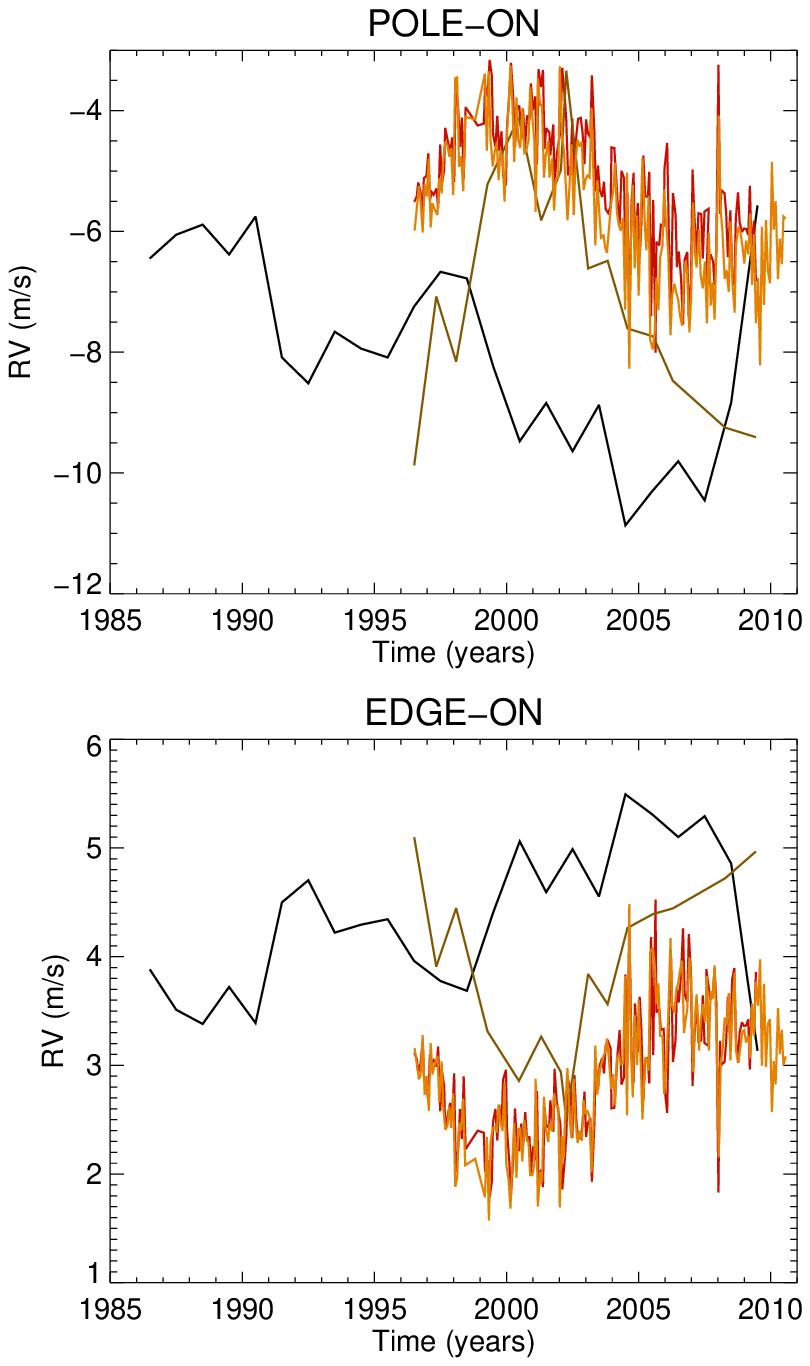}
\caption{
Time series of integrated meridional flows for pole-on (upper panel) and edge-on configuration (lower panel), for the U10 reconstruction (black), BA10 (brown), HR10 (res), and HR11 (orange). 
}
\label{temps2}
\end{figure}

Figure~\ref{temps2} compares the time series of integrated meridional flows for U10, B10, HR10, and HR11. The amplitudes are of the same order of magnitude, but B10, HR10, and HR11 are more anti-correlated with the cycle phase than U10. This may be due to the fact that they do not trace the same layers.


\section{Uncertainties on integrated RV}

We computed uncertainties on each value of mc(t) based on two different assumptions. First, we consider that for a given latitudinal profile (one of those shown in Fig.~\ref{lat}), the uncertainties between latitudinal bins are uncorrelated: we compute synthetic latitudinal profiles which are the sum of the original one added to a random deviation derived from a Gaussian distribution corresponding to the uncertainty in this latitude bin. We perform 25 such realisations of the time series: the rms of the RV values after integration at each time step is then the uncertainty for a given time step. The resulting uncertainties are very small, of the order of 0.06 m/s for pole-on time series and 0.02 m/s for the edge-on one. 

However, we cannot exclude the fact that part of the uncertainties on the latitudinal profiles are correlated, and correspond to an uncertainty on the global amplitude of the MC. Therefore, we also consider the possibility that all these uncertainties are correlated, i.e. that the uncertainty is mostly due to a bias on the global amplitude of the latitudinal profile, to provide an upper limit. Here we compute two integrated RV values, one for the upper envelope of the latitudinal profile, and the other one for the lower envelope, which provides an upper limit of the uncertainty on the integrated flow at each time step, shown in Fig.~\ref{temps1}:  the typical uncertainties on integrated RV are of the order of 1 m/s for the pole-on time series and 0.5 m/s for the edge-on one. Synthetic time series of mc(t) assuming such uncertainties are then generated in order to compute uncertainties on other variables such as the rms, amplitude, or correlation (see Sect.~2.3.2). It is likely that the true values lie between these two estimates. 

\clearpage
\onecolumn

\begin{longtable}{llll}
\caption{ \label{tab_hr11} Integrated meridional flow from \cite{hathaway11} }\\
\hline
Time   & RV at 0$^\circ$  & RV at 55$^\circ$   & RV at 90$^\circ$  \\
(year) & (m/s)     & (m/s) & (m/s)   \\  
\hline
1996.52 &  -5.98 &  -0.03 &   3.16 \\
1996.68 &  -5.25 &  -0.07 &   2.87 \\
1996.75 &  -5.51 &  -0.03 &   2.91 \\
1996.82 &  -6.01 &  -0.08 &   3.28 \\
1996.90 &  -5.22 &  -0.02 &   2.73 \\
1996.97 &  -5.42 &  -0.05 &   2.90 \\
1997.06 &  -4.80 &  -0.05 &   2.58 \\
1997.13 &  -5.93 &  -0.06 &   3.20 \\
1997.20 &  -5.53 &  -0.03 &   2.92 \\
1997.28 &  -5.64 &  -0.05 &   3.01 \\
1997.38 &  -5.72 &  -0.06 &   3.09 \\
1997.45 &  -5.12 &  -0.01 &   2.66 \\
1997.55 &  -5.37 &  -0.12 &   3.04 \\
1997.61 &  -4.84 &  -0.05 &   2.62 \\
1997.68 &  -4.51 &  -0.04 &   2.41 \\
1997.76 &  -4.58 &  -0.04 &   2.45 \\
1997.84 &  -4.76 &  -0.07 &   2.62 \\
1997.90 &  -4.62 &  -0.02 &   2.43 \\
1997.98 &  -5.39 &   0.01 &   2.74 \\
1998.06 &  -3.46 &  -0.04 &   1.88 \\
1998.13 &  -3.45 &  -0.08 &   1.96 \\
1998.20 &  -4.93 &   0.03 &   2.47 \\
1998.28 &  -4.76 &   0.02 &   2.40 \\
1998.36 &  -5.33 &   0.02 &   2.69 \\
1998.43 &  -4.10 &   0.01 &   2.08 \\
1998.82 &  -4.15 &  -0.00 &   2.14 \\
1999.18 &  -3.39 &  -0.02 &   1.79 \\
1999.26 &  -4.66 &   0.03 &   2.34 \\
1999.34 &  -3.35 &   0.07 &   1.57 \\
1999.39 &  -4.46 &   0.00 &   2.29 \\
1999.48 &  -4.92 &   0.05 &   2.42 \\
1999.55 &  -4.67 &  -0.01 &   2.43 \\
1999.63 &  -4.54 &  -0.02 &   2.38 \\
1999.70 &  -5.15 &   0.01 &   2.64 \\
1999.77 &  -4.66 &  -0.04 &   2.48 \\
1999.85 &  -4.80 &   0.03 &   2.40 \\
1999.93 &  -5.51 &  -0.00 &   2.84 \\
1999.99 &  -4.58 &   0.11 &   2.09 \\
2000.07 &  -3.95 &   0.04 &   1.95 \\
2000.15 &  -3.25 &  -0.00 &   1.68 \\
2000.22 &  -3.90 &  -0.05 &   2.12 \\
2000.29 &  -4.85 &   0.01 &   2.47 \\
2000.38 &  -3.78 &  -0.01 &   1.97 \\
2000.45 &  -4.48 &  -0.05 &   2.41 \\
2000.52 &  -4.62 &  -0.02 &   2.42 \\
2000.60 &  -3.96 &  -0.03 &   2.12 \\
2000.68 &  -4.72 &  -0.03 &   2.50 \\
2000.74 &  -4.67 &   0.02 &   2.36 \\
2000.84 &  -4.48 &   0.01 &   2.28 \\
2000.89 &  -3.61 &  -0.06 &   1.99 \\
2000.98 &  -4.12 &   0.06 &   1.99 \\
2001.05 &  -4.50 &   0.10 &   2.09 \\
2001.11 &  -5.67 &   0.02 &   2.88 \\
2001.18 &  -3.47 &   0.04 &   1.70 \\
2001.28 &  -3.92 &   0.00 &   2.02 \\
2001.34 &  -3.89 &   0.03 &   1.92 \\
2001.41 &  -5.11 &  -0.03 &   2.71 \\
2001.48 &  -4.76 &   0.01 &   2.43 \\
2001.58 &  -4.50 &   0.01 &   2.30 \\
2001.62 &  -5.10 &   0.07 &   2.46 \\
2001.72 &  -4.79 &   0.03 &   2.39 \\
2001.80 &  -5.65 &   0.07 &   2.74 \\
2001.87 &  -5.14 &   0.00 &   2.64 \\
2001.96 &  -5.09 &   0.01 &   2.61 \\
2002.01 &  -3.27 &  -0.01 &   1.70 \\
2002.09 &  -4.34 &  -0.05 &   2.35 \\
2002.16 &  -4.62 &   0.10 &   2.14 \\
2002.24 &  -5.17 &  -0.00 &   2.66 \\
2002.30 &  -5.81 &   0.01 &   2.97 \\
2002.38 &  -5.28 &   0.08 &   2.54 \\
2002.46 &  -4.56 &   0.01 &   2.33 \\
2002.55 &  -5.73 &   0.03 &   2.87 \\
2002.61 &  -4.84 &  -0.00 &   2.50 \\
2002.69 &  -4.48 &   0.01 &   2.29 \\
2002.77 &  -4.59 &  -0.01 &   2.39 \\
2002.85 &  -4.51 &   0.01 &   2.31 \\
2002.92 &  -5.31 &   0.02 &   2.68 \\
2002.99 &  -5.14 &   0.04 &   2.55 \\
2003.06 &  -5.08 &   0.01 &   2.58 \\
2003.14 &  -5.11 &   0.05 &   2.52 \\
2003.22 &  -3.95 &   0.01 &   2.02 \\
2003.29 &  -5.08 &   0.07 &   2.45 \\
2003.36 &  -6.23 &   0.01 &   3.18 \\
2003.43 &  -5.15 &  -0.03 &   2.73 \\
2003.52 &  -5.76 &   0.01 &   2.95 \\
2003.57 &  -6.10 &   0.06 &   3.00 \\
2003.67 &  -6.06 &   0.02 &   3.09 \\
2003.74 &  -6.27 &   0.04 &   3.13 \\
2003.80 &  -6.35 &   0.01 &   3.24 \\
2003.88 &  -5.90 &   0.01 &   3.03 \\
2003.96 &  -5.54 &   0.03 &   2.78 \\
2004.04 &  -5.33 &  -0.08 &   2.92 \\
2004.11 &  -4.87 &  -0.06 &   2.64 \\
2004.18 &  -5.68 &  -0.06 &   3.06 \\
2004.25 &  -5.89 &  -0.03 &   3.10 \\
2004.32 &  -5.98 &   0.03 &   3.01 \\
2004.41 &  -5.78 &  -0.02 &   3.02 \\
2004.49 &  -7.30 &  -0.03 &   3.82 \\
2004.56 &  -5.03 &   0.02 &   2.54 \\
2004.64 &  -8.27 &  -0.10 &   4.48 \\
2004.71 &  -6.13 &   0.02 &   3.12 \\
2004.78 &  -5.23 &  -0.01 &   2.71 \\
2004.85 &  -6.61 &  -0.03 &   3.47 \\
2004.94 &  -5.74 &  -0.02 &   3.01 \\
2005.00 &  -7.06 &   0.01 &   3.60 \\
2005.07 &  -5.93 &  -0.01 &   3.07 \\
2005.14 &  -4.78 &  -0.02 &   2.50 \\
2005.24 &  -5.95 &   0.00 &   3.06 \\
2005.30 &  -6.31 &   0.08 &   3.06 \\
2005.38 &  -5.41 &  -0.07 &   2.95 \\
2005.44 &  -7.80 &  -0.01 &   4.04 \\
2005.53 &  -7.95 &   0.01 &   4.08 \\
2005.61 &  -7.66 &   0.01 &   3.92 \\
2005.67 &  -6.76 &  -0.03 &   3.56 \\
2005.75 &  -7.30 &   0.00 &   3.75 \\
2005.82 &  -6.43 &   0.02 &   3.27 \\
2005.90 &  -6.22 &  -0.03 &   3.27 \\
2005.97 &  -5.30 &  -0.00 &   2.74 \\
2006.04 &  -5.86 &   0.02 &   2.98 \\
2006.12 &  -6.37 &  -0.07 &   3.45 \\
2006.20 &  -7.72 &  -0.08 &   4.17 \\
2006.26 &  -7.07 &   0.04 &   3.55 \\
2006.35 &  -6.87 &   0.02 &   3.49 \\
2006.42 &  -7.00 &  -0.03 &   3.68 \\
2006.50 &  -7.15 &  -0.04 &   3.77 \\
2006.57 &  -7.42 &  -0.03 &   3.88 \\
2006.64 &  -7.56 &  -0.03 &   3.96 \\
2006.72 &  -6.71 &   0.07 &   3.30 \\
2006.78 &  -6.49 &  -0.03 &   3.42 \\
2006.84 &  -6.68 &  -0.03 &   3.50 \\
2006.92 &  -7.67 &  -0.02 &   3.99 \\
2006.99 &  -7.11 &  -0.04 &   3.76 \\
2007.06 &  -5.52 &  -0.04 &   2.94 \\
2007.13 &  -6.37 &  -0.08 &   3.45 \\
2007.20 &  -6.59 &  -0.11 &   3.65 \\
2007.28 &  -5.97 &  -0.03 &   3.14 \\
2007.34 &  -6.72 &   0.02 &   3.40 \\
2007.43 &  -6.94 &   0.01 &   3.56 \\
2007.51 &  -6.50 &  -0.02 &   3.40 \\
2007.59 &  -7.51 &   0.02 &   3.82 \\
2007.66 &  -6.02 &  -0.01 &   3.14 \\
2007.72 &  -7.23 &   0.02 &   3.68 \\
2007.80 &  -6.84 &  -0.04 &   3.61 \\
2007.88 &  -7.46 &  -0.03 &   3.92 \\
2007.96 &  -5.97 &  -0.03 &   3.15 \\
2008.03 &  -4.08 &  -0.02 &   2.16 \\
2008.11 &  -5.65 &  -0.05 &   3.03 \\
2008.18 &  -6.01 &  -0.01 &   3.12 \\
2008.26 &  -6.78 &   0.02 &   3.44 \\
2008.33 &  -6.64 &   0.02 &   3.37 \\
2008.41 &  -7.00 &  -0.03 &   3.66 \\
2008.47 &  -5.83 &  -0.02 &   3.05 \\
2008.55 &  -6.89 &   0.00 &   3.54 \\
2008.64 &  -7.34 &  -0.03 &   3.85 \\
2008.68 &  -6.22 &   0.03 &   3.13 \\
2008.78 &  -5.92 &   0.01 &   3.03 \\
2008.84 &  -5.99 &  -0.03 &   3.15 \\
2008.93 &  -6.23 &  -0.04 &   3.31 \\
2009.00 &  -6.53 &   0.02 &   3.32 \\
2009.07 &  -6.15 &  -0.02 &   3.21 \\
2009.15 &  -6.38 &   0.01 &   3.27 \\
2009.22 &  -5.71 &  -0.06 &   3.07 \\
2009.30 &  -6.88 &  -0.01 &   3.57 \\
2009.36 &  -6.05 &  -0.07 &   3.28 \\
2009.44 &  -7.52 &   0.05 &   3.77 \\
2009.52 &  -6.77 &  -0.03 &   3.55 \\
2009.59 &  -8.21 &   0.11 &   3.97 \\
2009.66 &  -5.92 &   0.02 &   3.00 \\
2009.74 &  -7.22 &  -0.01 &   3.75 \\
2009.82 &  -5.81 &   0.00 &   2.99 \\
2009.89 &  -6.64 &   0.02 &   3.38 \\
2009.97 &  -6.85 &   0.05 &   3.42 \\
2010.05 &  -4.85 &  -0.03 &   2.57 \\
2010.11 &  -5.97 &   0.02 &   3.04 \\
2010.18 &  -5.50 &   0.00 &   2.83 \\
2010.28 &  -6.79 &  -0.01 &   3.53 \\
2010.34 &  -6.12 &   0.02 &   3.11 \\
2010.42 &  -6.54 &   0.02 &   3.33 \\
2010.48 &  -5.75 &  -0.02 &   3.00 \\
2010.57 &  -5.79 &  -0.04 &   3.08 \\
\hline
\end{longtable}

\end{appendix}

\end{document}